\documentclass[conference]{IEEEtran}
\usepackage{fancyhdr} 
\IEEEoverridecommandlockouts
\usepackage[x11names]{xcolor,colortbl}
\usepackage{makecell}

\usepackage{graphicx}
\usepackage{subcaption}
\usepackage{amsmath}
\usepackage{amsfonts}
\usepackage{amssymb}
\usepackage{url}
\usepackage{enumitem}
\usepackage{booktabs}
\usepackage[group-separator={,},group-minimum-digits={3}]{siunitx}
\usepackage[hidelinks]{hyperref}

\usepackage{sc24repro}

\usetag{explanation}
\usetag{example}

\usepackage[most]{tcolorbox}

\tcbset{width=\columnwidth,boxrule=0pt,colback=red,arc=0pt,auto outer arc,left=0pt,right=0pt,boxsep=5pt}

\definecolor{customcolor}{HTML}{9494b8}

\newtcolorbox[auto counter]{ObservationBox}{
    borderline west={3pt}{0pt}{DeepSkyBlue4},
    colback=customcolor!30!white}

\newcommand{\obs}[1]{\begin{ObservationBox} \textbf{Observation~\thetcbcounter:} #1 \end{ObservationBox}}

\def\BibTeX{{\rm B\kern-.05em{\sc i\kern-.025em b}\kern-.08em
    T\kern-.1667em\lower.7ex\hbox{E}\kern-.125emX}}

\usepackage{multirow}
\usepackage{tabularx}
\newcolumntype{b}{>{\scriptsize}X}
\newcolumntype{e}{>{\scriptsize}p}

\definecolor{r1}{RGB}{87,114,158}
\definecolor{r2}{RGB}{204,137,99}
\definecolor{r3}{RGB}{93,157,107}
\definecolor{r4}{RGB}{196,78,82}
\definecolor{r5}{RGB}{129,114,180}
\usepackage{marginnote}
\usepackage{soul} %
\usepackage{hhline}
\definecolor{lightyellow}{RGB}{250, 250, 180}
\definecolor{HLYELLOW}{RGB}{240, 127, 0}
\definecolor{hlyellow}{RGB}{240, 127, 0}
\sethlcolor{lightyellow}
\definecolor{lightcyan}{RGB}{160,255,255}
\definecolor{lightgreen}{rgb}{0.56, 0.93, 0.56}
\usepackage{xparse}

\usepackage{mdframed}
\global\mdfdefinestyle{review}{%
linecolor=lightyellow,linewidth=3pt,%
leftmargin=0cm,rightmargin=0cm,%
skipabove=0cm,skipbelow=0cm,%
innerrightmargin=0cm,innerleftmargin=0cm,%
innerbottommargin=0cm,innertopmargin=0cm,%
backgroundcolor=lightyellow
}
\global\mdfdefinestyle{reviewtext}{%
linecolor=lightyellow,linewidth=0pt,%
leftmargin=0cm,rightmargin=0cm,%
skipabove=0.1cm,skipbelow=0.1cm,%
innerrightmargin=0cm,innerleftmargin=0cm,%
innerbottommargin=0cm,innertopmargin=0cm,%
backgroundcolor=lightyellow
}

\global\mdfdefinestyle{shepherd}{%
linecolor=lightgreen,linewidth=3pt,%
leftmargin=0cm,rightmargin=0cm,%
skipabove=0cm,skipbelow=0cm,%
innerrightmargin=0cm,innerleftmargin=0cm,%
innerbottommargin=0cm,innertopmargin=0cm,%
backgroundcolor=lightgreen
}
\global\mdfdefinestyle{shepherdtext}{%
linecolor=lightyellow,linewidth=0pt,%
leftmargin=0cm,rightmargin=0cm,%
skipabove=0.1cm,skipbelow=0.1cm,%
innerrightmargin=0cm,innerleftmargin=0cm,%
innerbottommargin=0cm,innertopmargin=0cm,%
backgroundcolor=lightgreen
}

\begin{document}

\DeclareDocumentCommand\review{m g g}{%
    {\IfNoValueF {#2}{%
    \IfNoValueF {#3}{%
    {\marginnote{\sethlcolor{#3}\hl{\normalfont \textbf{{\normalsize{\color{white}#2}}}}}%
    }%
    }%
    \IfNoValueT {#3}{%
    {\marginnote{\normalfont \textbf{\normalsize{#2}}}%
    }%
    }%
    }%
    \hl{#1}%
    }%
}
\title{Exploring GPU-to-GPU Communication:\\Insights into Supercomputer Interconnects}

\makeatletter %
\newcommand{\linebreakand}{%
  \end{@IEEEauthorhalign}
  \hfill\mbox{}\par
  \mbox{}\hfill\begin{@IEEEauthorhalign}
}
\makeatother %

\author{\IEEEauthorblockN{Daniele De Sensi}
\IEEEauthorblockA{\textit{Sapienza University of Rome} \\
\textit{desensi@di.uniroma1.it} \\
}
\and
\IEEEauthorblockN{Lorenzo Pichetti}
\IEEEauthorblockA{\textit{University of Trento}\\
\textit{lorenzo.pichetti@unitn.it} \\
}
\and
\IEEEauthorblockN{Flavio Vella}
\IEEEauthorblockA{\textit{University of Trento}\\
\textit{flavio.vella@unitn.it} \\
}
\and
\IEEEauthorblockN{Tiziano De Matteis}
\IEEEauthorblockA{\textit{Vrije Universiteit Amsterdam}\\
\textit{t.de.matteis@vu.nl} \\
}
\and

\linebreakand 

\IEEEauthorblockN{Zebin Ren}
\IEEEauthorblockA{\textit{Vrije Universiteit Amsterdam}\\
\textit{z.ren@vu.nl} \\
}
\and
\IEEEauthorblockN{Luigi Fusco}
\IEEEauthorblockA{\textit{ETH Zurich}\\
\textit{luigi.fusco@inf.ethz.ch} \\
}
\and
\IEEEauthorblockN{Matteo Turisini}
\IEEEauthorblockA{\textit{CINECA}\\
\textit{m.turisini@cineca.it} \\
}
\and
\IEEEauthorblockN{Daniele Cesarini}
\IEEEauthorblockA{\textit{CINECA}\\
\textit{d.cesarini@cineca.it} \\
}
\and
\IEEEauthorblockN{Kurt Lust}
\IEEEauthorblockA{\textit{University of Antwerp}\\
\textit{kurt.lust@uantwerpen.be} \\
}
\and

\linebreakand 

\IEEEauthorblockN{Animesh Trivedi$^{*}$\thanks{$^{*}$Work done while at Vrije Universiteit Amsterdam.}}
\IEEEauthorblockA{\textit{IBM Research Europe}\\
\textit{animesh.trivedi@ibm.com} \\
}
\and
\IEEEauthorblockN{Duncan Roweth}
\IEEEauthorblockA{\textit{HPE Cray}\\
\textit{duncan.roweth@hpe.com} \\
}
\and
\IEEEauthorblockN{Filippo Spiga}
\IEEEauthorblockA{\textit{NVIDIA}\\
\textit{fspiga@nvidia.com} \\
}
\and
\IEEEauthorblockN{Salvatore Di Girolamo}
\IEEEauthorblockA{\textit{NVIDIA}\\
\textit{sdigirolamo@nvidia.com} \\
}
\and
\IEEEauthorblockN{Torsten Hoefler}
\IEEEauthorblockA{\textit{ETH Zurich}\\
\textit{htor@inf.ethz.ch} \\
}
}

\IEEEaftertitletext{\vspace{-3\baselineskip}}

\maketitle
\thispagestyle{fancy}
\lhead{}
\rhead{}
\chead{}
\lfoot{\footnotesize{
SC24, November 17-22, 2024, Atlanta, Georgia, USA
\newline 979-8-3503-5291-7/24/\$31.00 \copyright 2024 IEEE}}
\rfoot{}
\cfoot{}
\renewcommand{\headrulewidth}{0pt}
\renewcommand{\footrulewidth}{0pt}

\begin{abstract}
Multi-GPU nodes are increasingly common in the rapidly evolving landscape of exascale supercomputers. On these systems, GPUs on the same node are connected through dedicated networks, with bandwidths up to a few terabits per second. However, gauging performance expectations and maximizing system efficiency is challenging due to different technologies, design options, and software layers. This paper comprehensively characterizes three supercomputers — Alps, Leonardo, and LUMI — each with a unique architecture and design. We focus on performance evaluation of intra-node and inter-node interconnects on up to \num{4096} GPUs, using a mix of intra-node and inter-node benchmarks. By analyzing its limitations and opportunities, we aim to offer practical guidance to researchers, system architects, and software developers dealing with multi-GPU supercomputing. Our results show that there is untapped bandwidth, and there are still many opportunities for optimization, ranging from network to software optimization.
\end{abstract}

\section{Introduction}

Supercomputers are a key infrastructure enabling advancements in several science domains and transformative societal changes. New workloads' computing requirements, ranging from machine learning (ML) to scientific computing and extending to big-data analytics, are driving supercomputer architecture evolution. Due to their massive parallelism, energy efficiency, and memory bandwidth, GPUs became the core of such evolution, characterized by the development of multi-GPU nodes and high-performance intra-node interconnection networks. Nowadays, exascale~\cite{10.1145/3372390} and pre-exascale systems in the Top500~\cite{top500} are equipped with up to 8 GPUs per node, connected with fast dedicated networks with bandwidth up to 3.6 Tb/s per direction~\cite{nvidiaNVLinkNVSwitch}. At the same time, due to a steady increase in computing and memory requirements, the number of nodes increased to tens of thousands of nodes~\cite{9658212}, leading to systems with up to \num{75000} GPUs~\cite{10.1145/3581784.3607089,metacluster}.

Moving data efficiently across such a high number of GPUs is challenging for multiple reasons. First, there is a significant interconnect, topology, and hardware diversity, thus making the mapping of communications to the underlying system non-trivial. Secondly, on the software side, programmers can rely on different software solutions, ranging from manually copying data between GPU memory on a single node to using transparent and higher-level GPU-Aware solutions such as MPI~\cite{9059272} or NCCL/RCCL~\cite{nccl-1,rccl-1} (referred as \textit{*CCL} in the rest of the paper) across nodes. However, the most effective approach for managing a large number of GPUs and the maturity level of the software is still unclear. Lastly, large-scale network-related effects such as congestion and \textit{network noise}~\cite{gpcnet,staci2018,Desensi2020} can severely impact the scalability when increasing the number of GPUs, thus hampering the computational power and high bandwidth of these systems.

To investigate the aforementioned challenges, we comprehensively characterize \emph{three supercomputers with different architectures}: Alps (NVIDIA H100 GPUs and HPE Cray Slingshot interconnect~\cite{Desensi2020}), Leonardo (NVIDIA A100 GPUs and NVIDIA InfiniBand HDR interconnect), and LUMI (AMD Instinct\textsuperscript{TM} MI250X GPUs and HPE Cray Slingshot interconnect). We systematically benchmark the performance of intra-node GPU networks (Sec.~\ref{sec:intra-node} and Sec.~\ref{sec:intra-node-coll}), and inter-node networks (Sec.~\ref{sec:inter-node} and Sec.~\ref{sec:congestion}) up to \num{4096} GPUs on the three supercomputers. Our study includes a detailed analysis of data movement performed through explicit device-to-device copies, *CCL, and GPU-Aware MPI. Lastly, we evaluate the impact of network noise on GPU-GPU data movements (Sec.~\ref{sec:congestion}), showing that it can severely impact workload scalability.

This paper provides the first at-scale study characterizing multi-GPU interconnect performance across hardware technologies and diverse communication APIs and software stacks. We spotlight several sources of inefficiencies, ranging from routing to communication libraries tuning. We show that the best way to move data between GPUs depends on several factors like transfer size, communication pattern, and number of GPUs, and might change across systems. We present eight key observations, offering valuable insights to system architects, researchers, practitioners, and software developers to optimize data movements in large-scale multi-GPU systems and exploit current and upcoming systems to their fullest.

\begin{table*}
\let\center\empty
\let\endcenter\relax
\centering

\begin{tabularx}{\linewidth}{ e{2.5cm} e{4cm} b b }
\\[0.5ex]

& {\centering \textbf{Alps (\#6 in Top500)}}     & {\centering\textbf{Leonardo (\#7 in Top500)}}             & {\centering\textbf{LUMI (\#5 in Top500)}} \\[1.0ex] 
\hline
\rowcolor{lightgray!20} CPU  & 72-core NVIDIA \textit{Grace}  & 32-core Intel \textit{Ice Lake} Xeon 8358 & 64-core AMD \textit{Trento} EPYC 7A53 \\[1.0ex] 

GPU & 4x NVIDIA Hopper H100 & 4x NVIDIA Ampere A100  (special SKU) & 4x AMD MI250X (8 GCDs) \\[1.0ex] 

\rowcolor{lightgray!20} NICs & 4x HPE Cray 200 Gb/s Cassini-1 & 2x dual-port NVIDIA Connect-X6 (100 Gb/s per port) (i.e., 4x 100 Gb/s ports) & 4x HPE Cray 200 Gb/s Cassini-1 \\[1.0ex] 

Intra-node Interconnect & NVLink 4.0, 6x links towards any other GPU (1.2 Tb/s between any pair) & NVLink 3.0, 4x links towards any other GPU (800 Gb/s between any pair)  & Between one and four 400 Gb/s Infinity Fabric links towards other GCDs (see Fig.~\ref{fig:lumi}) \\[1.0ex] 

\rowcolor{lightgray!20} Inter-node Interconnect & HPE Cray Slingshot 11. Dragonfly topology & NVIDIA Infiniband HDR. Dragonfly+ topology with 23 groups. Each group is a 2-level \textit{fat-tree} &  HPE Cray Slingshot 11. Dragonfly topology with 24 groups\\[1.0ex] 

Software Environment & Cray MPICH \texttt{v8.1.28}, libfabric \texttt{v1.15.2}, CUDA \texttt{v12.3}, \texttt{aws-ofi-nccl} plugin  & Open MPI \texttt{v4.1.4} (relying on UCX \texttt{v1.13.0}), CUDA \texttt{v12.1} &  Cray clang \texttt{v16.0.1}, Cray MPICH \texttt{v8.1.27}, libfabric \texttt{v1.15.2}, ROCM \texttt{v5.7.1.1}, \texttt{aws-ofi-rccl} plugin (\texttt{v1.4})  \\[1.0ex] 
 \hline
\end{tabularx}
\caption{Main characteristics of the analyzed systems. Top500 rankings from June 2024 list.}
\label{tab:systems}
\end{table*}

\section{Systems Description}\label{sec:systems}
In the following, we describe the main characteristics of the three analyzed systems, and we summarize them in Table~\ref{tab:systems}. Since all the analyzed interconnects are full-duplex, \textit{when we refer to intra-node and inter-node network bandwidth, we will always report the \emph{unidirectional} bandwidth and express it in bits per second for consistency}. %

\subsection{Alps}\label{sec:systems:alps}
Alps is a 270 PFlop/s supercomputer ranked 6th in the Top500 (June 2024). It is deployed by CSCS~\cite{alps} and currently under provisioning. Thus, some presented results might be subject to further tuning in the upcoming months before opening to production. For this paper, we used the early access Santis partition~\cite{santis}, where we had access to 512 nodes. %

\textbf{Node architecture} Each node is composed of four GH200 \textit{Grace Hopper Superchip}~\cite{gh200doc} connected in an all-to-all topology using NVLink 4.0. Six 200 Gb/s links connect each GH200 pair, for a total of 1.2 Tb/s between any GPU pair (see Fig.~\ref{fig:alps}). Every GH200 has 96 GB HBM3 and 120 GB LPDDRX5 memory. Every node acts as a single NUMA system (composed of 8 NUMA domains), with 288 CPU cores and 4 GPUs. 

\textbf{Inter-node connectivity} Each node has one HPE Cray Cassini-1 200 Gb/s Network Interface Card (NIC) for each GH200. Nodes are connected in a Dragonfly topology~\cite{dragonfly} through an HPE Cray Slingshot-11 network~\cite{Desensi2020,khorassani2023high}. Each switch has 16 ports to endpoints, 31 ports to other switches within the same Dragonfly group, and 17 ports to switches in other Dragonfly groups.

\subsection{Leonardo}\label{sec:systems:leonardo}
\emph{Leonardo}~\cite{turisini2023leonardo} is a 240 PFlop/s supercomputer, ranked 7th on the Top500~\cite{top500} (June 2024). It is owned by the EuroHPC Joint Undertaking and hosted by CINECA. We consider Leonardo's \emph{Booster} GPU partition, consisting of \num{3456} computing nodes.

\textbf{Node architecture} Each node is equipped with a single socket 32-core Intel Xeon\textsuperscript{\textregistered} 8358 CPU and four NVIDIA A100 \emph{TensorCore} GPUs~\cite{a100tcdoc} (\num{13824} GPUs in total).
Each node has 512 GB CPU memory, organized in eight DDR4 slots, and 64 GB HBM2e memory per GPU. %
Within a node, GPUs are connected through NVIDIA NVLink 3.0, with each GPU connected to each of the other three GPUs through four 200 Gb/s links (see Fig.~\ref{fig:leonardo}). The intra-node communication is completed by a 256 Gb/s 16-lane PCIe\textsuperscript{\textregistered} Gen4.0 bus per GPU, used to communicate with the host CPU and with the NIC. 
\begin{figure}[htpb]
\centering
\begin{subfigure}{0.45\linewidth}
    \includegraphics[width=\columnwidth]{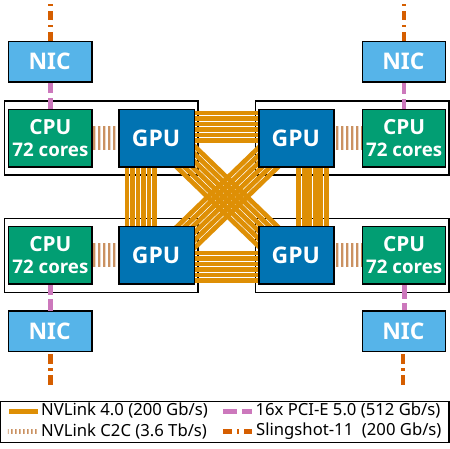}
    \caption{Alps.}
    \label{fig:alps}
\end{subfigure}
\hfill
\begin{subfigure}{0.45\linewidth}
    \includegraphics[width=\columnwidth]{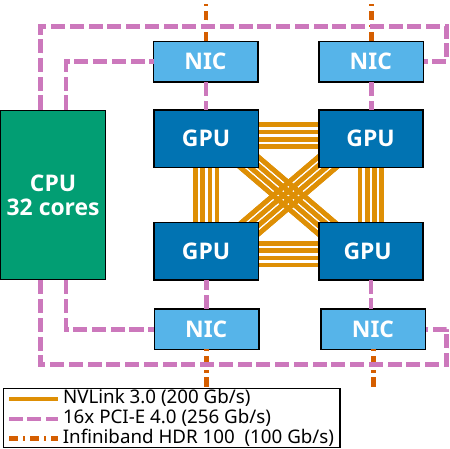}
    \caption{Leonardo.}
    \label{fig:leonardo}
\end{subfigure}
    \caption{Alps and Leonardo (Booster) node architectures.}
    \label{fig:alps:leo}
\end{figure}

\textbf{Inter-node connectivity} Nodes are interconnected through an InfiniBand HDR network, and each node is equipped with two 200 Gb/s dual port NVIDIA Connect-X6 NICs. Each node thus has four 100 Gb/s network ports, all connected to the same switch at the time of writing. For this paper, we consider them as being four separate NICs. Nodes are connected through a Dragonfly+~\cite{dragonflyplus}, with each group containing 180 nodes and structured as a two-level fat tree. Each group has 18 spine and 18 leaf switches. Switches have 40 200Gb/s ports (each of which can be configured as 2 100Gb/s ports). Leaf switches connect 40 100Gb/s ports to 10 nodes (4 GPUs per node) and 18 200 Gb/s ports to spine switches (with 2 200Gb/s ports unused). Spine switches connect 18 200Gb/s ports to leaf switches and 22 200Gb/s ports to other spine switches in different Dragonfly+ groups.

\begin{figure}
    \centering
    \includegraphics[width=0.8\columnwidth]{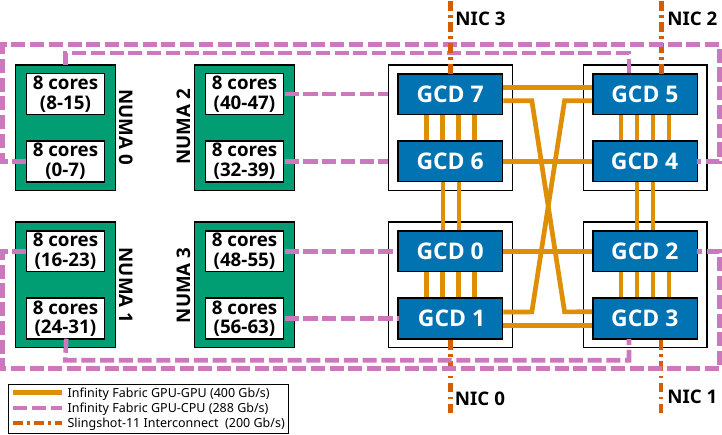}
    \caption{LUMI-G node architecture.}
    \label{fig:lumi}
\end{figure}
\subsection{LUMI}\label{sec:systems:lumi}
LUMI is a 380 PFlop/s supercomputer ranked 5th in the Top500 (June 2024). It is owned by the EuroHPC Joint Undertaking and hosted by CSC~\cite{lumisupercomputerLUMIsFull}. This paper considers the \textit{LUMI-G} GPU partition consisting of \num{2978} nodes.

\textbf{Node architecture} Each node has one 64-core AMD EPYC\textsuperscript{TM} 7A53 ``Trento'' CPU, configured as 4 NUMA domains, each accessing 128 GB of DDR4 memory. 
Each node also contains 4 AMD MI250X GPUs~\cite{amdcdna}, each with 2 Graphics Compute Dies~(GCDs), for a total of 8 GCDs per node. Each die 
has access to a 64 GB slice of HBM memory for a total of 
128 GB memory per MI250x module. Because one module is seen as two separate GPUs from a software perspective, \emph{for the rest of this paper, we consider a LUMI node as an 8 GPU node}. Each GCD is connected to a NUMA node through a 288 Gb/s AMD Infinity Fabric\textsuperscript{TM} (IF) link. GCDs are connected to each other with one to four 400 Gb/s IF links, as shown in Fig.~\ref{fig:lumi}. %

\textbf{Inter-node connectivity} Each MI250X module is connected to a 200 GiB/s Cassini-1 NIC. Nodes are connected in a Dragonfly topology through an HPE Cray Slingshot-11 interconnect composed of 24 groups, with 124 nodes per group. 
Each node is connected to two different switches in the same group. Each switch has 16 ports to endpoints, 31 ports to other switches within the same Dragonfly group, and 17 ports to switches in other Dragonfly groups.

\section{Intra-Node Point-to-Point Performance}\label{sec:intra-node}
We start our analysis by assessing the performance of the intra-node GPU-GPU interconnection for point-to-point communications. We first describe the benchmarking methodology (Sec.~\ref{sec:intra-node-methodology}) and the performance tuning performed on each system (Sec.~\ref{sec:intra-node-tuning}). We then analyze the point-to-point performance (Sec.~\ref{sec:intra-node:pp}), focusing on LUMI intra-node architecture (Sec.~\ref{sec:intra-node:pp:lumi}).

\subsection{Benchmarking Methodology}\label{sec:intra-node-methodology}
In all the experiments we have a separate MPI process managing each GPU in the system. We set the affinity of the processes so that each MPI rank manages the GPU closest to the core it is mapped to. We run each experiment between \num{100} times and \num{1000} times (depending on the transfer size). For experiments involving collective communication, we report the maximum time (or minimum goodput) across all the participating ranks~\cite{10.1145/2807591.2807644}. We do not include communicators' creation time. 
Unless specified otherwise, we always refer to unidirectional bandwidth in Gb/s. We analyze the performance of point-to-point and collective communication using different mechanisms and techniques to transfer data between GPUs. Namely, we consider:
\begin{itemize}[leftmargin=*]
\item \textbf{Trivial Staging:} We copy buffers to and from the GPU memory to the host memory. Then, we transfer data between processes using MPI. This is a trivial implementation to use as a baseline. We pinned the memory but implemented no pipelining or parallel copies between memories. Data thus moves in a \textit{store-and-forward} fashion. For point-to-point transfers, we can estimate the peak goodput by summing the time required to transfer the data from device memory to host memory and the time to copy the data between two host memory buffers.
\item \textbf{Device-Device Copy:} Buffers are copied directly from the device to the device memory. We share memory handles across the processes managing the different GPUs, allowing them to transfer data directly between GPU memories. For the alltoall collective, each GPU copies data to all the other GPUs asynchronously to overlap the copies.
\item \textbf{*CCL:} We transfer data between GPUs using NCCL~\cite{nccl-1,nccl-2} (on Leonardo and Alps) or RCCL~\cite{rccl-1,rccl-2} (on LUMI).
\item \textbf{GPU-Aware MPI:} We transfer data using GPU-Aware MPI~\cite{6587715}.
\end{itemize}

We developed our benchmark from scratch due to some limitations of existing benchmarks. OSU~\cite{osu} lacks benchmarks for explicit device-device copy, and nccl-/rccl-tests~\cite{nccl-test,rccl-test} only support *CCL. Also, both do not report individual per-iteration timings, which are needed to assess network noise and performance variability (Sec.~\ref{sec:congestion}). Creating our benchmark ensured consistency across all communication mechanisms analyzed. We used \texttt{MPI\_Wtime} for timing individual iterations, with resolutions of 25ns on LUMI and Leonardo, and 30ns on Alps (measured experimentally). The timing excludes one-time operations like buffer allocation and handles exchange. The benchmark synchronizes with the GPU before stopping the timer to ensure full data receipt, except for MPI, where this is implicit. For *CCL, the timing includes the group start/end. This is consistent with nccl-/rccl-tests. We publicly released the code as part of the paper artifact.

\subsection{Performance Tuning}\label{sec:intra-node-tuning}
We tuned the performance of all analyzed systems, as the initial default configuration did not fully leverage their potential. This involved searching for and setting environment variables and analyzing their impact on performance. 

\textbf{*CCL} On Alps and LUMI we forced *CCL to ignore the CPU affinity set by Slurm (by setting the \texttt{NCCL\_IGNORE\_CPU\_AFFINITY} to 1), obtaining up to 1.6x performance improvement on alltoall and up to 6x on allreduce starting from two nodes. We further improved performance by 2x on alltoall and 3x on allreduce, by
increasing the distance at which the GPU can use Direct RDMA when communicating with the NIC (\texttt{NCCL\_NET\_GDR\_LEVEL=3}). On LUMI, we set \texttt{NCCL\_NCHANNELS\_PER\_PEER=32} for intra-node point-to-point tests, which led to a 3.5x performance improvement.

\begin{figure*}[htpb]
\centering
\begin{subfigure}{0.32\linewidth}
    \includegraphics[width=1\linewidth]{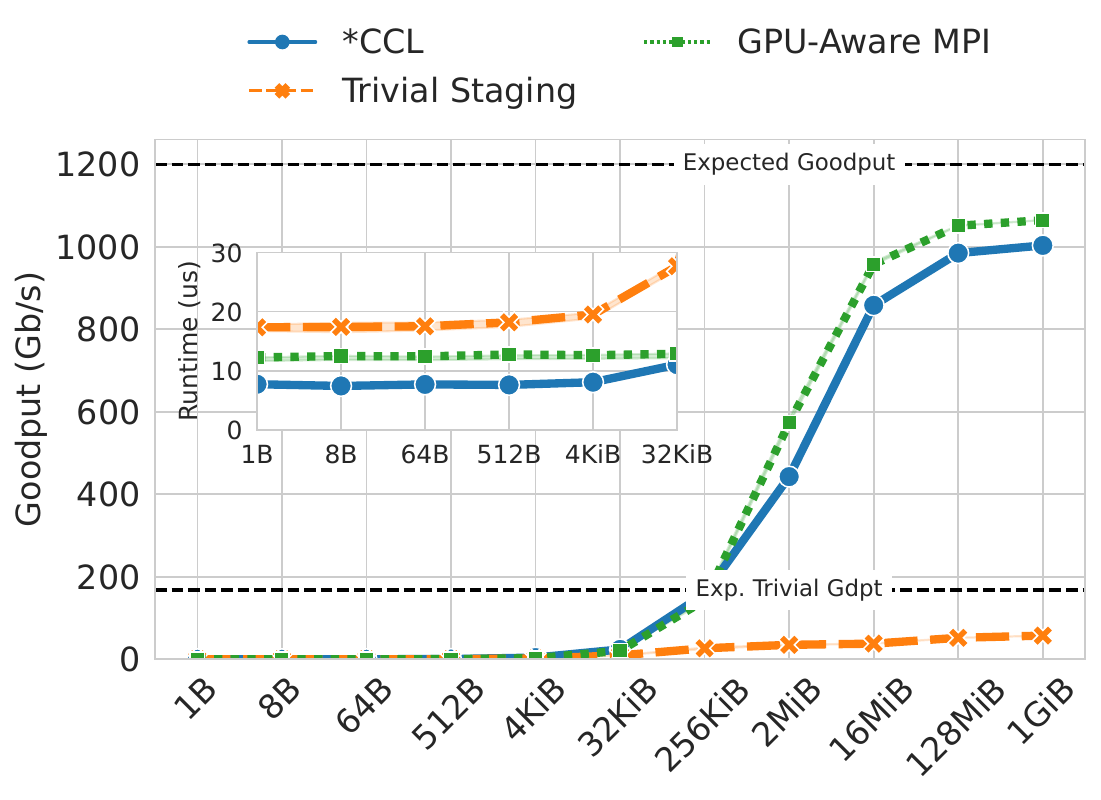}
    \caption{Alps}
    \label{fig:intra-node-pp-alps}
\end{subfigure}
\hfill
\begin{subfigure}{0.32\linewidth}
    \includegraphics[width=1\linewidth]{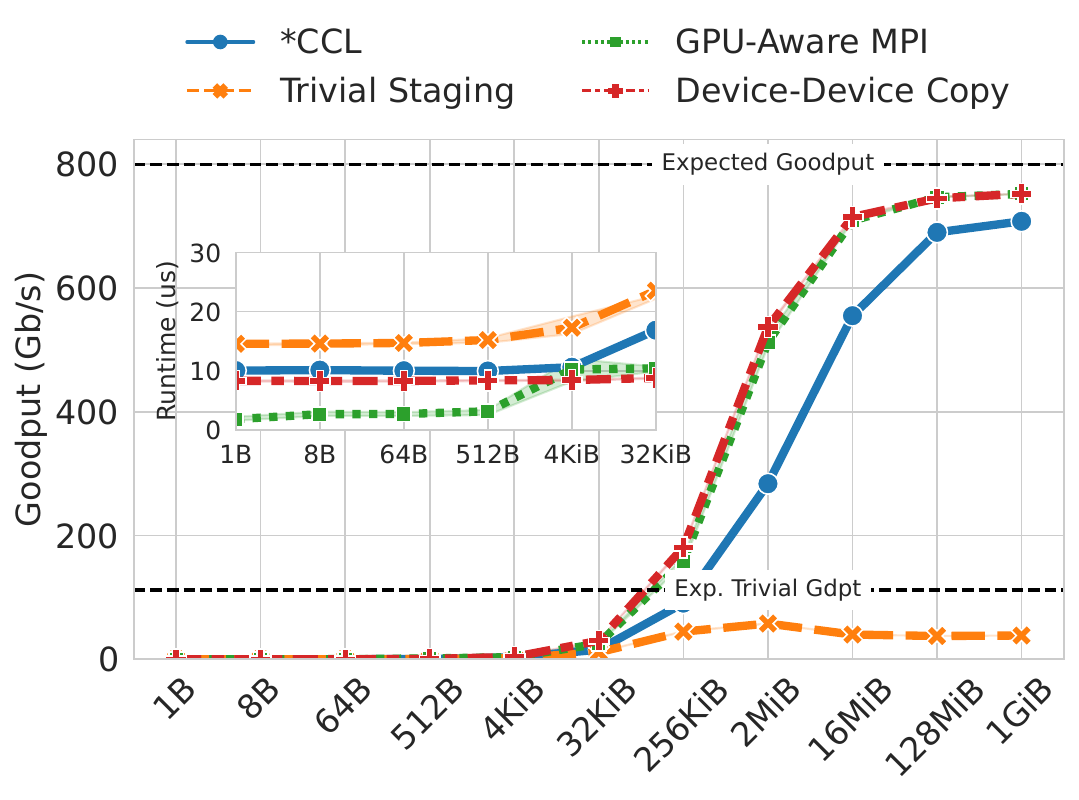}
    \caption{Leonardo}
    \label{fig:intra-node-pp-leonardo}
\end{subfigure}
\hfill
\begin{subfigure}{0.32\linewidth}
    \includegraphics[width=1\linewidth]{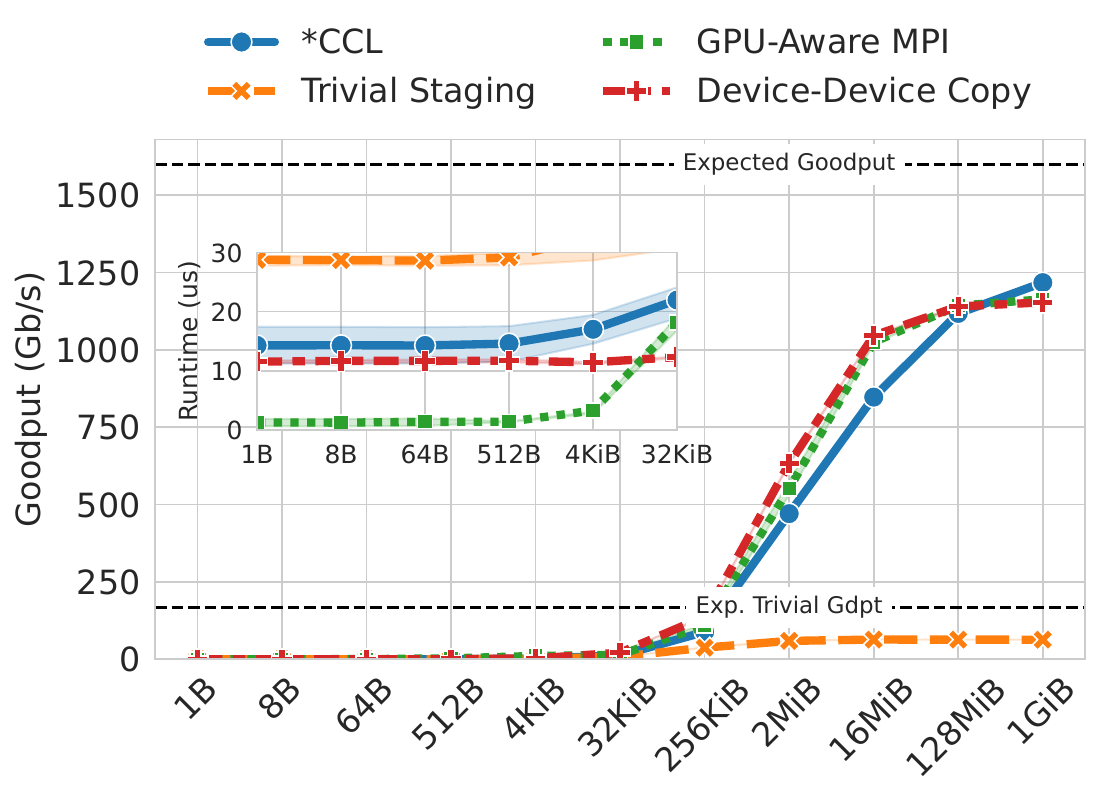}
    \caption{LUMI}
    \label{fig:intra-node-pp-lumi}
\end{subfigure}
\caption{GPU-GPU unidirectional transfers performance. For the sake of readability, we use different y-axis ranges.}
    \label{fig:intra-node-pp}
\end{figure*}

\textbf{GPU-Aware MPI} On Alps, to improve MPICH performance on small point-to-point transfers on a single node, we forced the use of device-device copies regardless of the transfer size (\texttt{MPICH\_GPU\_IPC\_THRESHOLD=1}), reducing runtime by 2x for transfers smaller than 4 KiB. We increased the size of the GPU-attached staging buffer used by MPICH for GPU-kernel-based optimizations for the allreduce to 128 MiB (\texttt{MPICH\_GPU\_ALLREDUCE\_BLK\_SIZE}), which led to a 50\% improvement on single-node allreduce. On LUMI, we disabled System Direct Memory Access (SDMA -- \texttt{HSA\_ENABLE\_SDMA=0}), increasing performance by up to 3x. On Leonardo, UCX was not loading GDRCopy~\cite{7116873} because it was installed in the wrong path. We fixed this by adding the correct path to \texttt{LD\_LIBRARY\_PATH}, increasing performance for small messages up to 6x.

The optimization of some of these parameters involved discussions with HPC site support teams and Cray/HPE, NVIDIA, and AMD engineers (e.g., *CCL parameters). Others were optimized by analyzing preliminary data. For instance, the runtime on Alps did not increase monotonically with message size, prompting further investigation and tuning of the IPC threshold. Understanding and resolving some of these unusual behaviors took several days of investigation.

\obs{Achieving good performance on multi-GPU systems requires non-trivial tuning, which depends on the system, message size, communication library, and number of nodes. The default choices made by *CCL and GPU-Aware MPI are not always optimal, and manual tuning can improve performance up to an order of magnitude.}

\subsection{Point-to-point Latency and Goodput}\label{sec:intra-node:pp}
In Fig.~\ref{fig:intra-node-pp}, we report the goodput between two GPUs on the same node, measured through a ping-pong microbenchmark for different transfer sizes. We report the unidirectional goodput (in Gb/s), defined as the number of bytes in the buffer divided by half the runtime. The inner plot reports the runtime (in microseconds) for small messages. Each data point represents the mean across the experiments, and the width of the shaded area around the line is the interquartile range (for some plots it is too small to be visible).  We report with dashed horizontal lines both the GPU-GPU unidirectional nominal goodput and the trivial staging expected goodput. For example, on Leonardo, any GPU pair is connected with four 200 Gb/s links (see Fig.~\ref{fig:leonardo}), for a total of 800 Gb/s nominal goodput per direction between any pair of GPUs. 

On LUMI, the peak GPU-GPU goodput depends on the specific GPU selected. For this experiment, we selected GPUs 0 and 1, connected through four 400 Gb/s Infinity Fabric links. In Sec.~\ref{sec:intra-node:pp:lumi}, we analyze the performance for different GPUs combinations. Moreover, disabling SDMA enables GPUs to use more than one Infinity Fabric link at a time~\cite{10.1145/3581784.3607089}. On Alps, we did not run the experiments involving explicit device to device copies since GPU peer access is not enabled on the nodes at the time being.

\textbf{Goodput} First, we observe that the goodput of trivial staging is up to one order of magnitude lower than the other implementations due to the low bandwidth when moving data between host and device memory. *CCL, MPI GPU-Aware, and device-device copies provide a comparable goodput.
On Leonardo, we observe a goodput for GPU-Aware on medium-sized messages that is up to 2x higher than that of NCCL. 

\textbf{Latency} When analyzing the runtime for small messages, we observe similar performance for *CCL and MPI on Alps, but a large performance gap on Leonardo and LUMI. On Leonardo, this is due to the use of GDRCopy~\cite{7116873}. On LUMI, Cray MPICH transfers small buffers between GPUs on the same node by copying them through host memory (rather than doing device-device copy), using an optimized \texttt{memcpy}, where the CPU issues load/store operations directly to GPU HBM. In contrast, on NVIDIA GPUs, CPU load/store operations to GPU memory are not permitted, resulting in higher latency on Alps.

\obs{GPU-Aware MPI provides the highest goodput for intra-node point-to-point transfers on all the analyzed systems. For small transfers, the optimal solution changes across the systems, depending on architectural features and specific optimization implemented by MPI.}

\subsection{Impact of GPU Location on LUMI}\label{sec:intra-node:pp:lumi}
On LUMI, each node has 8 GPUs, connected to each other with a different number of Infinity Fabric links, ranging from one to four (see Sec.~\ref{sec:systems:lumi}). We report in Fig.~\ref{fig:lumi-gpu-loc} the unidirectional goodput between GPU 0 and the other seven GPUs on a node when transferring a 1 GiB buffer. We denote with a dashed horizontal line the nominal goodput for each GPU pair, computed by considering the single path with the highest bandwidth between the two GPUs.

\begin{figure}[htpb]
    \centering
    \includegraphics[width=1\linewidth]{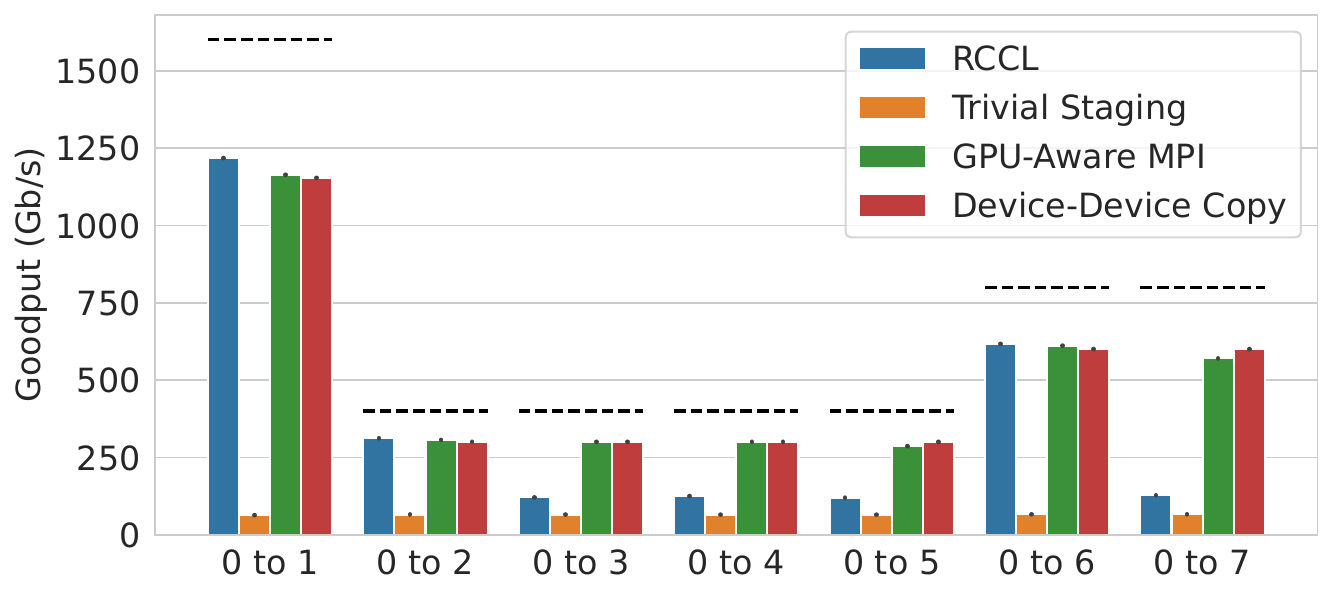}
    \caption{Unidirectional goodput from GPU 0 on LUMI to other GPUs on the same node, for a 1 GiB buffer. }
    \label{fig:lumi-gpu-loc}
\end{figure}

\begin{figure*}[htpb]
\centering
\begin{subfigure}{0.32\linewidth}
    \includegraphics[width=1\linewidth]{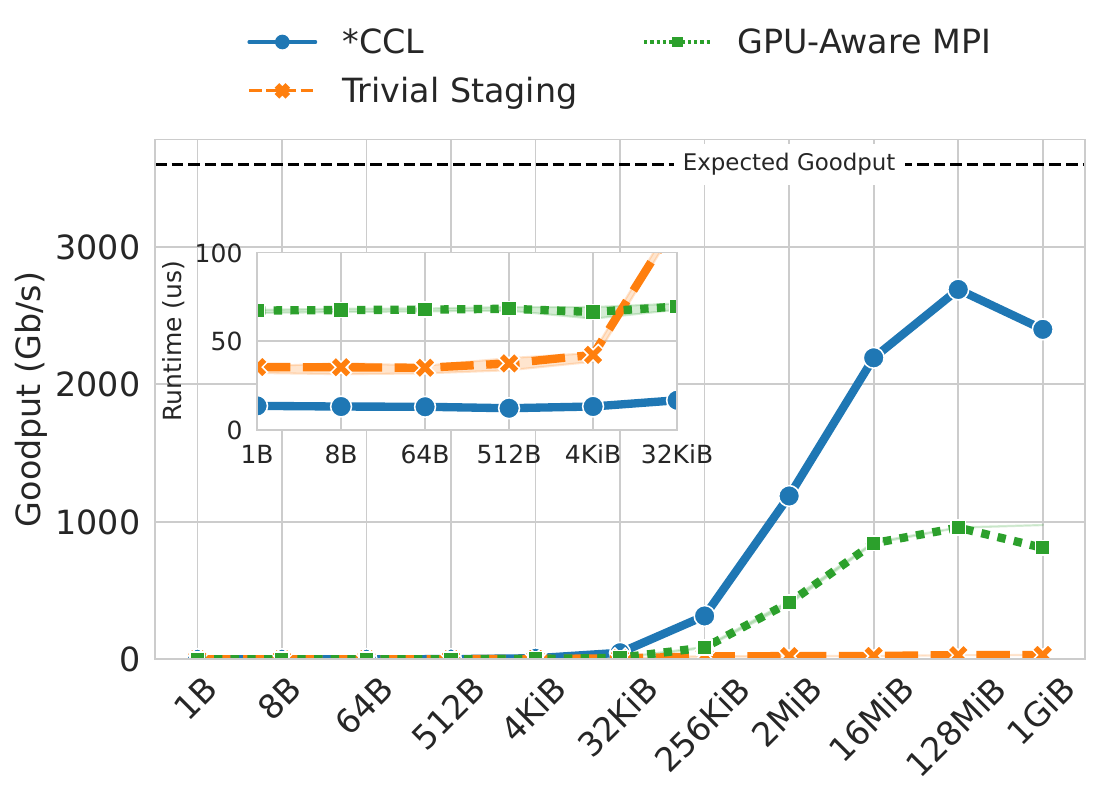}
    \caption{Alps}
    \label{fig:intra-node-a2a-alps}
\end{subfigure}
\hfill
\begin{subfigure}{0.32\linewidth}
    \includegraphics[width=1\linewidth]{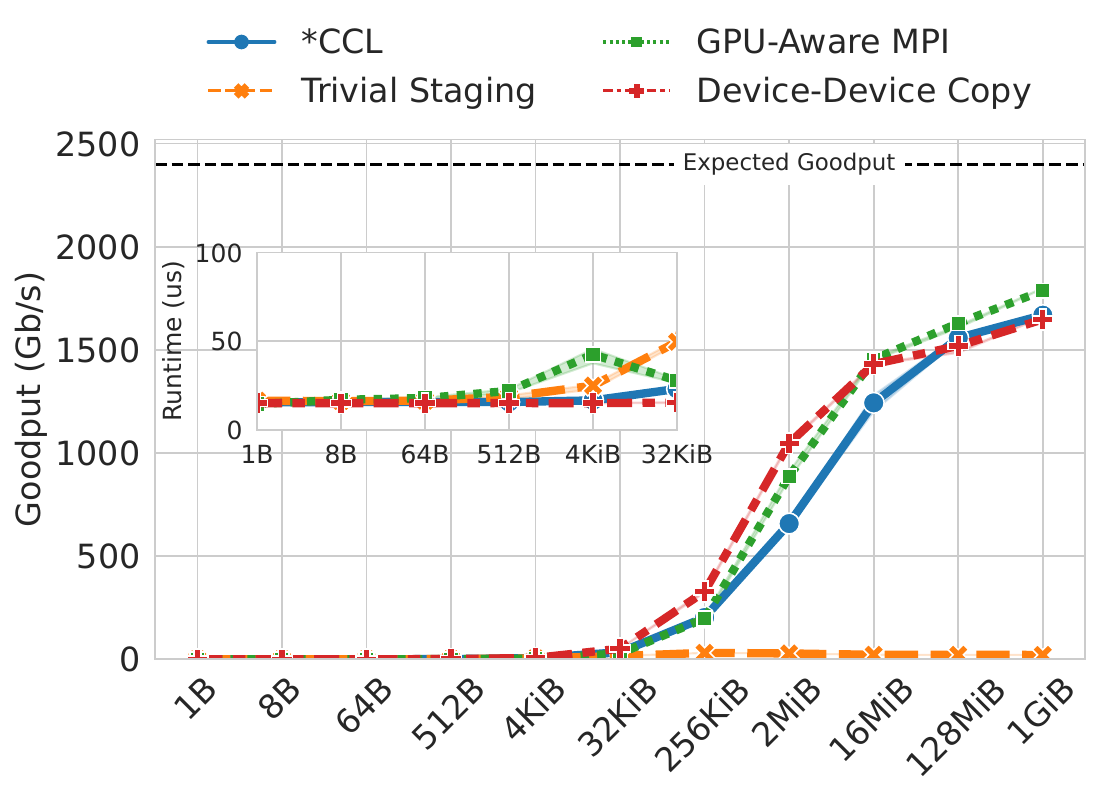}
    \caption{Leonardo}
    \label{fig:intra-node-a2a-leonardo}
\end{subfigure}
\hfill
\begin{subfigure}{0.32\linewidth}
    \includegraphics[width=1\linewidth]{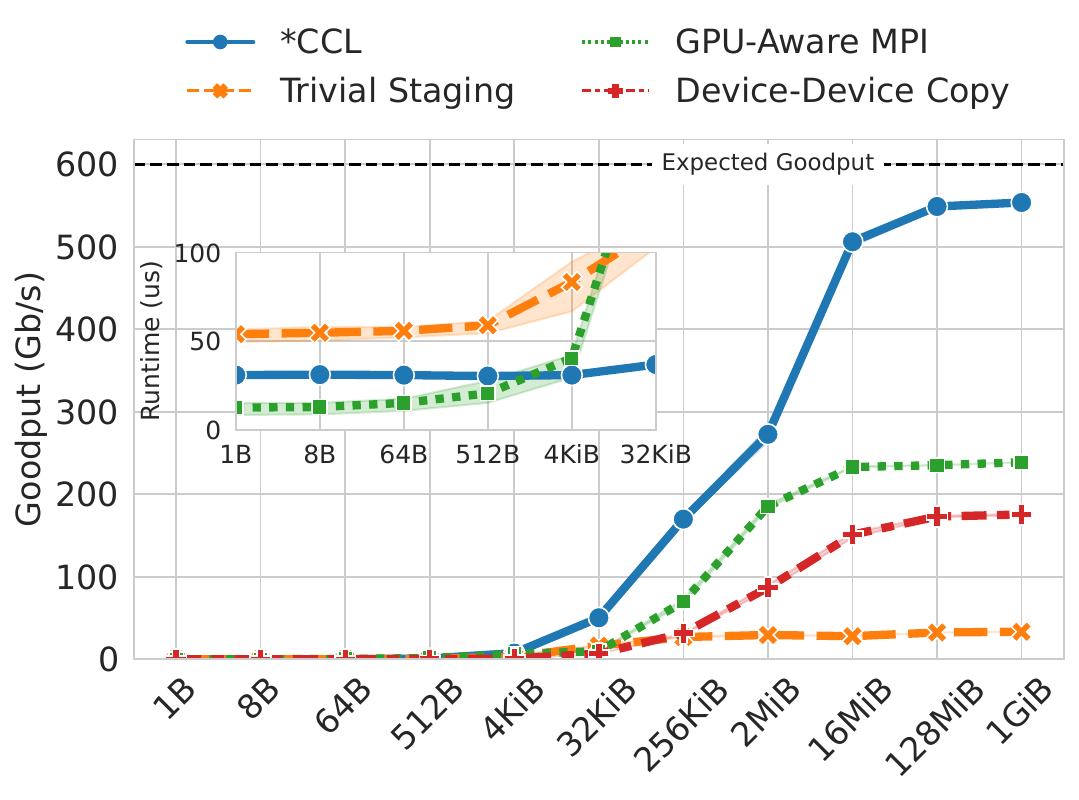}
    \caption{LUMI}
    \label{fig:intra-node-a2a-lumi}
\end{subfigure}
\caption{Intra-node alltoall performance. For the sake of readability, we use different y-axis ranges.}
    \label{fig:intra-node-a2a}
\end{figure*}

\begin{figure*}[htpb]
\centering
\begin{subfigure}{0.32\linewidth}
    \includegraphics[width=1\linewidth]{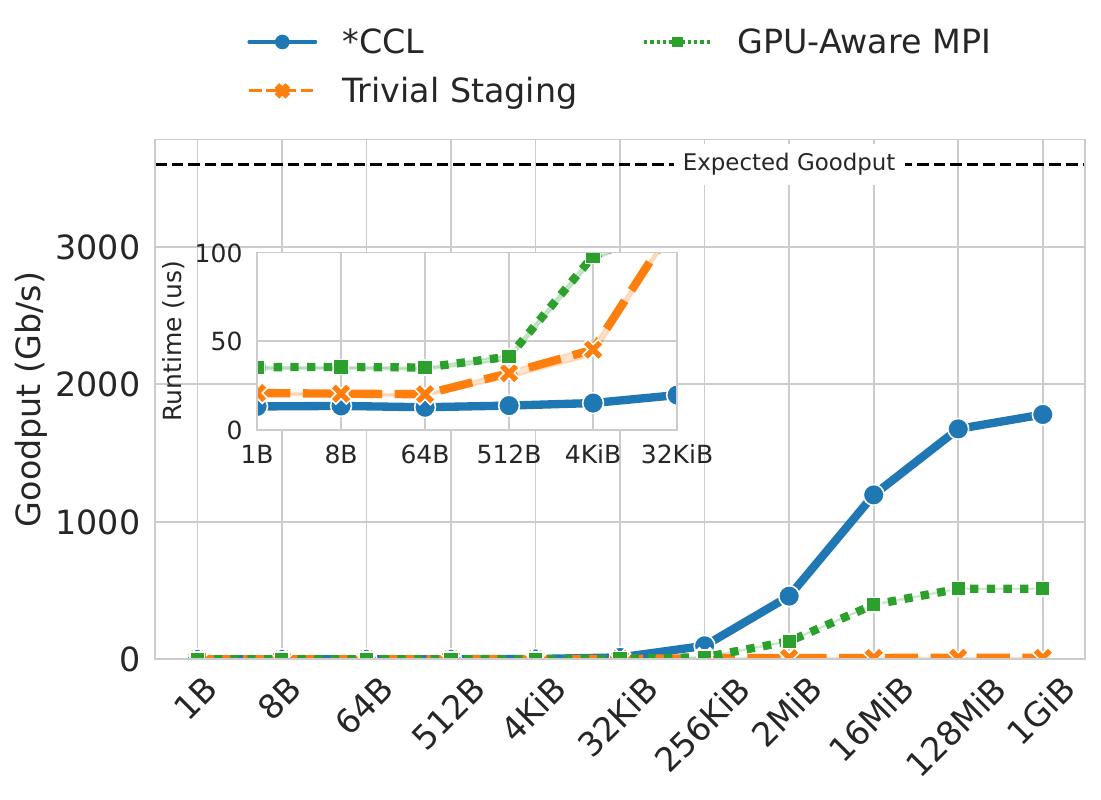}
    \caption{Alps}
    \label{fig:intra-node-ar-alps}
\end{subfigure}
\hfill
\begin{subfigure}{0.32\linewidth}
    \includegraphics[width=1\linewidth]{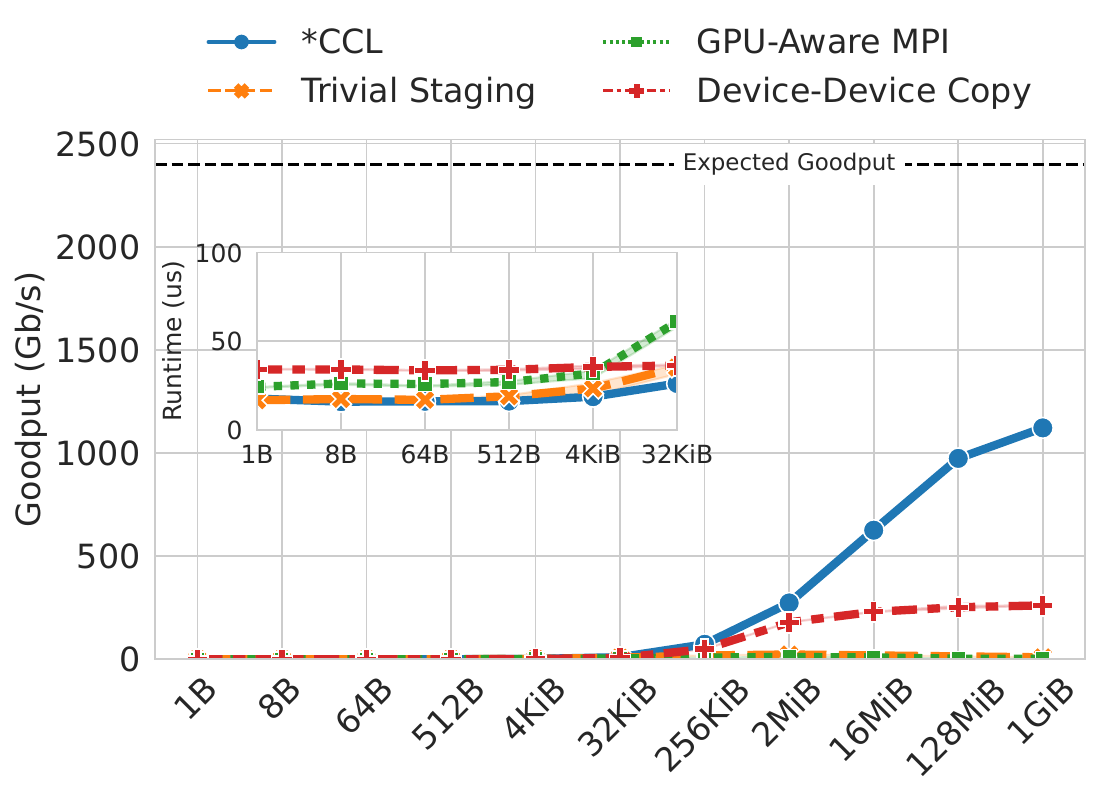}
    \caption{Leonardo}
    \label{fig:intra-node-ar-leonardo}
\end{subfigure}
\hfill
\begin{subfigure}{0.32\linewidth}
    \includegraphics[width=1\linewidth]{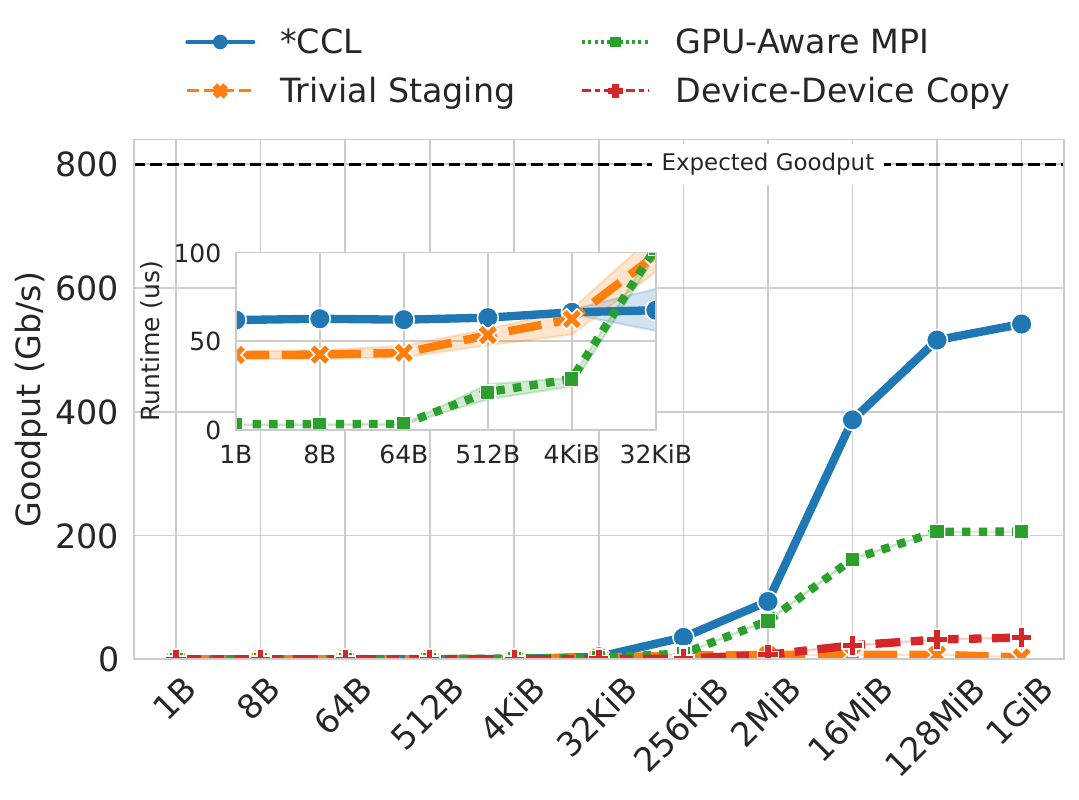}
    \caption{LUMI}
    \label{fig:intra-node-ar-lumi}
\end{subfigure}
\caption{Intra-node allreduce performance. For the sake of readability, we use different y-axis ranges.}
    \label{fig:intra-node-ar}
\end{figure*}

As expected, we do not observe any difference for the trivial staging, since data is not moved directly between GPUs but across the host memory. Both GPU-Aware MPI and the device-device copy achieve around 70\% of the nominal goodput on any GPU pair. On the other hand, in some cases (e.g., when GPU 0 and 5 communicate) RCCL achieves less than half the goodput of GPU-Aware MPI and device-device copy. By analyzing the data more in detail, we can observe that, although GPU-Aware MPI and device-device copy achieve the same goodput both towards GPU 6 and 7 (towards which GPU 0 has the same nominal goodput), RCCL achieves a much higher goodput towards GPU 6 compared to GPU 7. 

By analyzing RCCL debug information\footnote{This can be done by setting the environment variables \texttt{NCCL\_DEBUG\_SUBSYS=INIT,GRAPH} and \texttt{NCCL\_DEBUG=INFO}.}, we identified it assumes a lower available bandwidth towards GPU 7 compared to 6, and thus does not fully exploit the available bandwidth. We believe this might happen because the available bandwidth is estimated considering the number of hops rather than the number of paths connecting the two GPUs, which might be helpful for collective communications, where multiple GPUs concurrently communicate, sharing the available links. However, this setup underutilizes the GPU-GPU interconnect for sparser communication patterns. 

\obs{On LUMI, RCCL point-to-point communication primitives do not correctly determine the bandwidth available between GPUs on the same node, thus underutilizing the available bandwidth.}

\section{Intra-Node Collectives Performance}\label{sec:intra-node-coll}
We now focus on intra-node collective performance, by analyzing expected and measured goodput for alltoall (Sec.~\ref{sec:intra-node-coll:a2a:exp} and Sec.~\ref{sec:intra-node-coll:a2a:measured}) and allreduce (Sec.~\ref{sec:intra-node-coll:ar:exp} and Sec.~\ref{sec:intra-node-coll:ar:measured}).

\subsection{Alltoall Expected Goodput}\label{sec:intra-node-coll:a2a:exp}
For collectives, we define the goodput as the buffer size divided by the runtime.
To compute the expected goodput, we determined the edge forwarding index~\cite{HEYDEMANN1989103} of the graph corresponding to the intra-node GPUs connectivity. The edge forwarding index is defined as the maximum number of paths crossing any edge and gives an estimate of the maximum load across any network link and, thus, of the worst-case peak bandwidth (e.g., for an alltoall). 
On Alps and Leonardo, GPUs are fully connected, and each link is crossed by only one path (i.e., the maximum edge forwarding index is one), and we expect a peak goodput equal to the GPU injection bandwidth. 

On LUMI, assuming data is routed between GPUs using shortest paths, the most loaded link is the one between GCD 1 and 5 (and that between GCD 7 and 3), which is used in four separate paths. Thus, because each IF link has a 400 Gb/s bandwidth, we can expect a 100 Gb/s peak goodput between any pair of GCDs during an alltoall. Because any GCD can send data on six different IF links simultaneously, we expect a peak alltoall goodput of 600 Gb/s. It is worth noting that each GCD on the MI250X GPUs has the same injection bandwidth as one A100 GPU. However, the per-GPU alltoall goodput on LUMI is lower because the GCDs on the same node are not fully connected and, thus, the graph describing the intra-node connectivity has a higher edge forwarding index.

\subsection{Alltoall Measured Goodput}\label{sec:intra-node-coll:a2a:measured}
We report in Fig.~\ref{fig:intra-node-a2a} the measured goodput for an alltoall between all GPUs on a node, and the expected goodput (denoted with a dashed horizontal line). While MPI and RCCL natively provide an alltoall implementation, NCCL does not, and we implemented the alltoall with a trivial algorithm where each GPU sends the data to all the other nodes at the same time (as suggested in the documentation~\cite{nccl-doc}). We used the same algorithm to implement the alltoall using device-device copies. Moreover, we observed no performance difference when comparing RCCL native and trivial alltoall implementations.

On Alps and LUMI, *CCL provides the best performance for large transfers since *CCL collectives are specifically optimized for the target systems. For example, communications are mapped according to the specific topology, and the number of in-flight data chunks during pipelined operations is tuned by considering the bandwidth available between GPU pairs. Such fine-grained optimizations are not performed by MPI, which thus does not entirely exploit the bandwidth of the intra-node GPU interconnect. On Leonardo, *CCL provides slightly lower performance than MPI. For small transfers, on Alps and Leonardo the performance of *CCL is comparable with that of MPI. On the other hand, on LUMI, for small transfers GPU-Aware MPI is up to 3x faster than *CCL, consistent with what we observed for point-to-point transfers. %

\subsection{Allreduce Expected Goodput}\label{sec:intra-node-coll:ar:exp}%
On Alps and Leonardo, since each GPU is directly connected to the other GPUs on the same node and can receive from all the other GPUs at the same time, the optimal allreduce algorithm would consist of a pipelined ternary tree reduce (with one of the GPUs as the root and the other three as leaves) followed by a ternary tree broadcast. We thus estimate the peak goodput as the sum of the bandwidth across all the outgoing links from a GPU. On LUMI, since the GPUs are not fully connected, the optimal algorithm for large messages would be the Rabenseifner algorithm, with a ring reduce-scatter followed by a ring allgather~\cite{rabenseifner}. The specific connectivity between the GPUs allows for four edge-disjoint bidirectional rings~\cite{amdcdna}, each using 400 Gb/s Infinity Fabric links. Because the Rabenseifner algorithm sends twice the number of bytes in the buffer, we can thus expect 800 Gb/s peak goodput.

\subsection{Allreduce Measured Goodput}\label{sec:intra-node-coll:ar:measured}
In Fig.~\ref{fig:intra-node-ar}, we show the allreduce performance for different message sizes. On Alps and Leonardo, *CCL outperforms MPI at any transfer size. On LUMI, on the other hand, GPU-Aware MPI is characterized by the lowest runtime for small transfers whereas, although *CCL performs best on large transfers, its performance is far from the expected peak. This is consistent with what we observed for point-to-point transfers. 

GPU-Aware MPI exhibits low performance on all the analyzed systems, and we observe a higher performance gap between *CCL and GPU-Aware MPI on the allreduce compared to the alltoall. Indeed, the allreduce involves data aggregation, which *CCL performs on the GPUs. Although MPICH also performs data aggregation on the GPUs, we believe that *CCL coordinates GPU execution better. On Leonardo, we note an even larger gap, since Open MPI  runs the allreduce on the host~\cite{openmpi_gpu_allreduce}, similarly to what we do in the baseline implementation. It is worth remarking that on Leonardo Open MPI does not support UCC~\cite{ucc}. The implementation relying on device-device copies performs a reduction on GPU 0, followed by a broadcast. We do not implement any form of pipelining, and we mostly use it for reference and to show that implementing efficient multi-GPU collectives is non-trivial.

Last, we observe a higher gap between measured and expected performance on collectives compared to point-to-point communications, showing there is still space for collective algorithms optimization. Measured goodput on LUMI gets closer to the expected one. Indeed, LUMI has a lower expected goodput, which is thus easier to saturate.

\obs{For single node collectives, *CCL outperforms GPU-Aware MPI in most cases, except for small collectives on LUMI. Indeed, unlike MPI, *CCL collectives are optimized for the specific GPU models. Nevertheless, there is still room for collective algorithms optimization.}

\begin{figure*}[htpb]
\centering
\begin{subfigure}{0.32\linewidth}
    \includegraphics[width=1\linewidth]{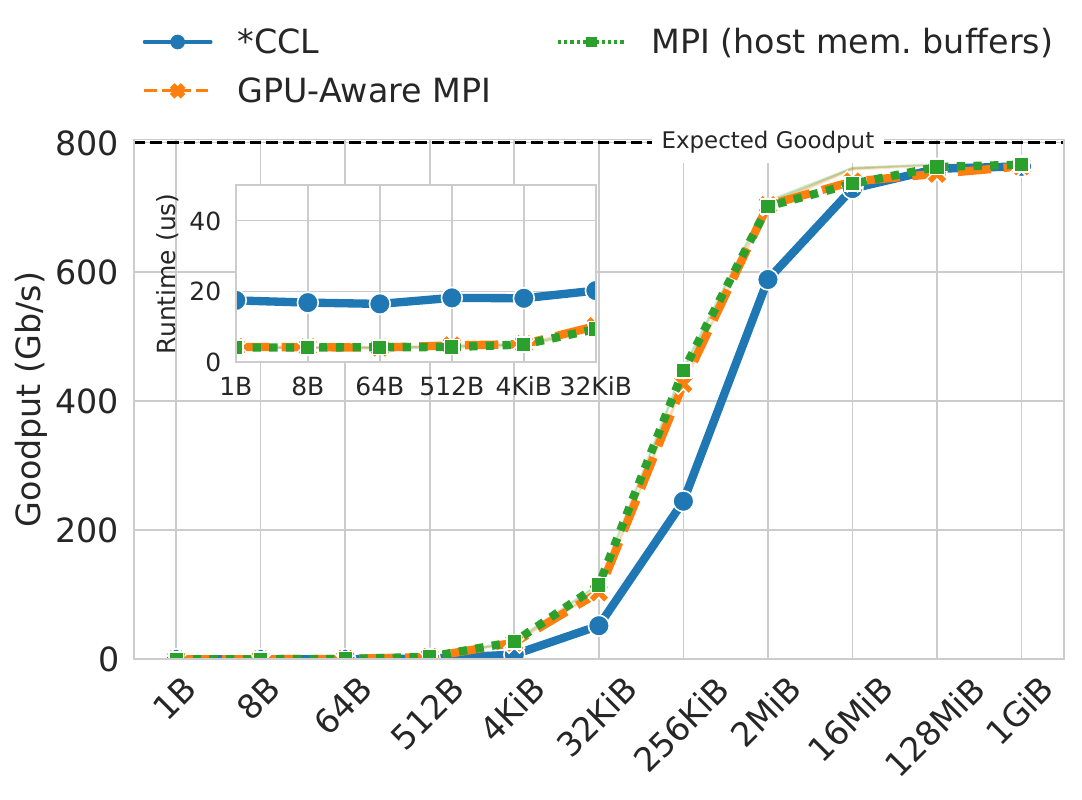}
    \caption{Alps}
    \label{fig:uni_bw:alps}
\end{subfigure}
\hfill
\begin{subfigure}{0.32\linewidth}
    \includegraphics[width=1\linewidth]{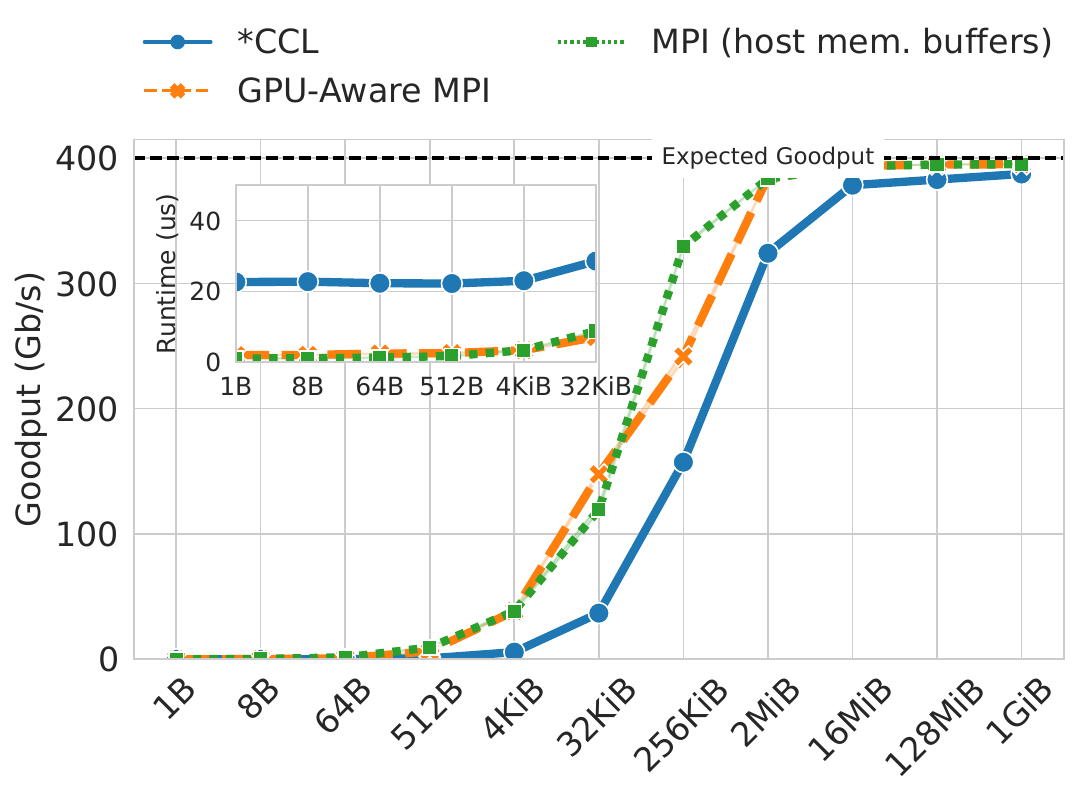}
    \caption{Leonardo}
    \label{fig:uni_bw:leonardo}
\end{subfigure}
\hfill
\begin{subfigure}{0.32\linewidth}
    \includegraphics[width=1\linewidth]{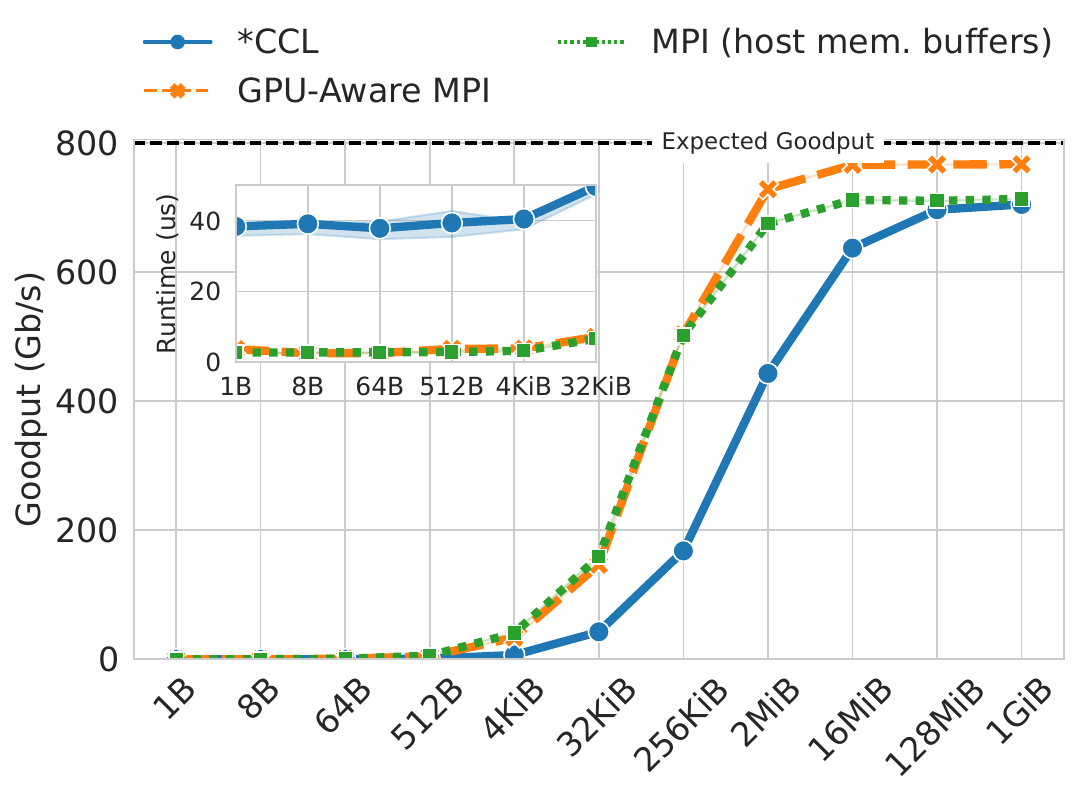}
    \caption{LUMI}
    \label{fig:uni_bw:lumi}
\end{subfigure}
\caption{Inter-node unidirectional goodput (per-node) and latency. For the sake of readability, we use different y-axis ranges.}
\label{fig:uni_bw}
\end{figure*}

\section{Inter-Node Performance}\label{sec:inter-node}
We now analyze the interconnect performance when running on multiple nodes. We first analyze point-to-point performance (Sec.~\ref{sec:inter-node:pp}), focusing on the impact of network distance on performance (Sec.~\ref{sec:inter-node:location}). Then, we analyze the performance of collective communications up to \num{4096} GPUs for alltoall (Sec.~\ref{sec:inter-node:coll:a2a}) and allreduce (Sec.~\ref{sec:inter-node:coll:ar}). We apply the benchmarking methodology described in Sec.~\ref{sec:intra-node-methodology}, and we run a number of MPI processes on each node equal to the number of available GPUs. For completeness, to assess the overhead of managing GPUs (especially for small transfers), we also analyze the performance when transferring buffers located in the host memory. In this case, we run one MPI process per NIC. On Alps and LUMI, we set the affinity of each process so that it uses the closest GPU and NIC. 

\subsection{Unidirectional Latency and Goodput}\label{sec:inter-node:pp}
We run a ping-pong test to measure the point-to-point goodput between two nodes, with each process exchanging data with the corresponding process on the other node.  We report the results of our analysis in Fig.~\ref{fig:uni_bw}. We report on the x-axis the number of bytes transmitted on each NIC, and, on the y-axis, the total goodput of the node (i.e., the sum of the goodput on each NIC). MPI provides the highest goodput and lowest latency on all the analyzed systems, regardless of whether the buffer is in host or GPU memory. This is mostly due to overheads introduced by *CCL when launching and managing GPU kernels.

\obs{On inter-node point-to-point communications, MPI outperforms *CCL by up to one order of magnitude on small transfers, and by up to 3x on larger transfers.}

\subsection{Impact of Network Distance on Performance}\label{sec:inter-node:location}
We now analyze the impact of the GPU network location on performance. Namely, on all systems, two GPUs on different nodes can be connected to the same switch, to two switches on the same Dragonfly/Dragonfly+ group, or to two switches in two different groups. We measure latency and goodput with a ping-pong test sending one byte and 1 GiB, respectively. We use MPI since we showed in Fig.~\ref{fig:uni_bw} that, for point-to-point transfers, it provides the highest bandwidth and lowest latency on all the analyzed systems. We report the results of our analysis in Fig.~\ref{fig:lat_bw_dist:gpu} for buffers allocated on GPU memory. For completeness and to separate the impact of the network from that of the GPU management, we report in Fig.~\ref{fig:lat_bw_dist:cpu} the same analysis, but for buffers allocated on host memory.

Each box's top and bottom borders represent the third and first quartiles, respectively. The middle line represents the median, the $\times$ marker the mean, and the top and bottom whiskers are the 5th and 95th percentile. Last, the notch around the median represents the 95\% confidence interval of the median. For the sake of readability, we do not report individual outliers but we annotate on the plot the minimum and maximum observed values.

\begin{figure*}[htpb]
\centering
\begin{subfigure}{0.49\linewidth}
    \includegraphics[width=1\linewidth]{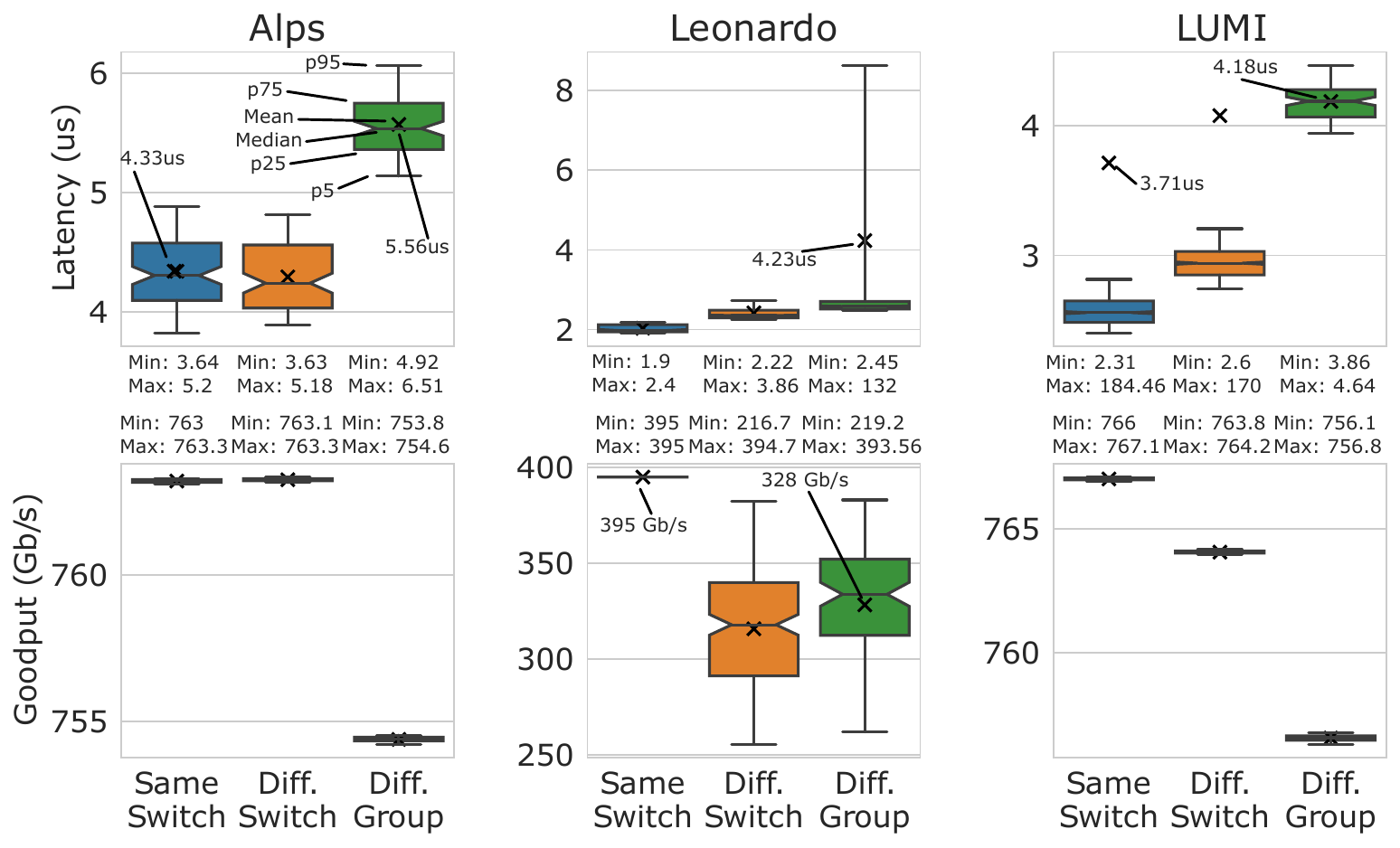}
    \caption{GPU memory buffers.}
    \label{fig:lat_bw_dist:gpu}
\end{subfigure}
\hfill
\begin{subfigure}{0.49\linewidth}
    \includegraphics[width=1\linewidth]{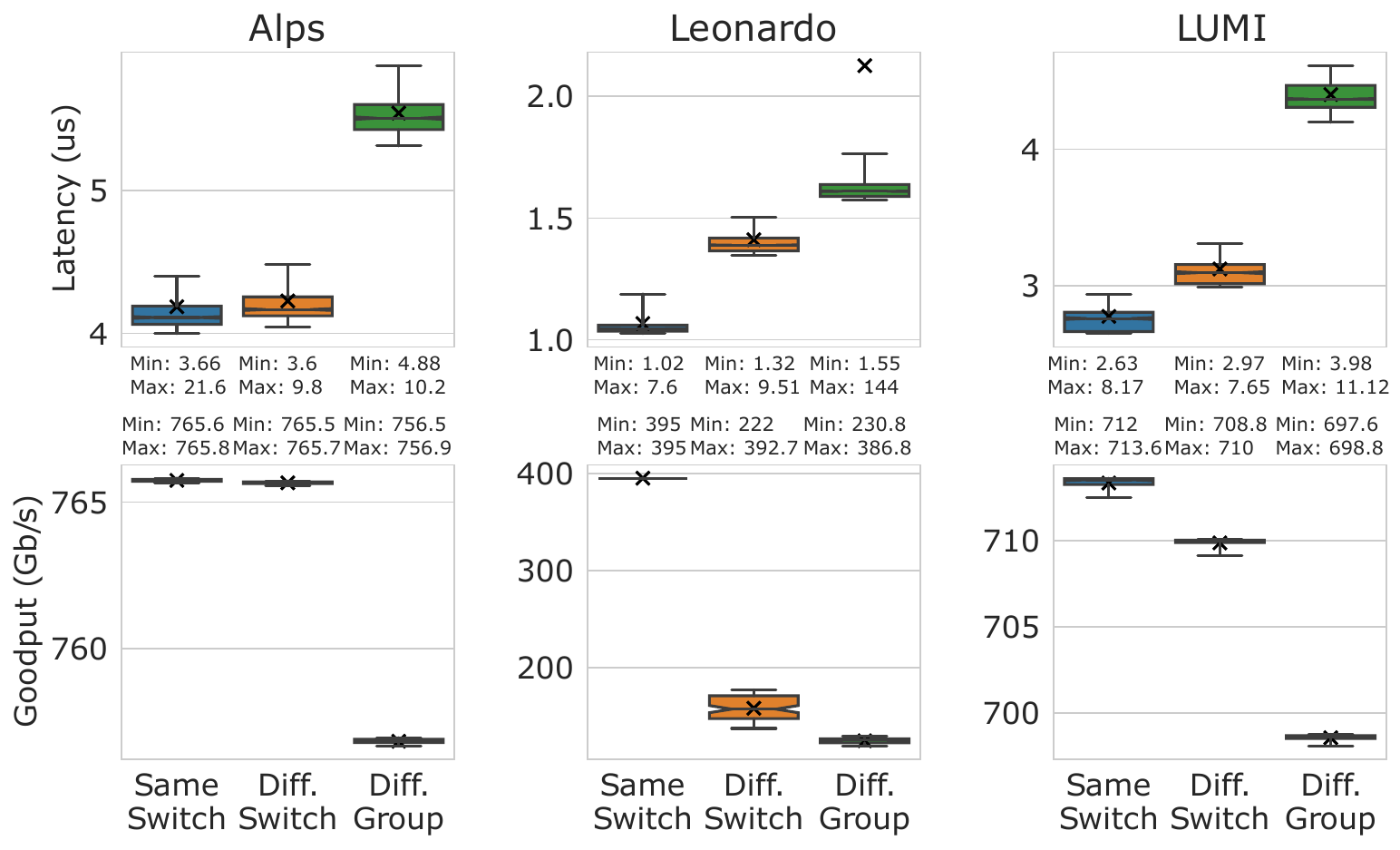}
    \caption{Host memory buffers.}
    \label{fig:lat_bw_dist:cpu}
\end{subfigure}
    \caption{Latency and goodput between GPUs at different distances. For the sake of readability, we use different y-axis ranges.}
\label{fig:lat_bw_dist}
\end{figure*}

\subsubsection{GPU memory buffers} We first focus on the performance when transferring buffers allocated on GPU memory (Fig.~\ref{fig:lat_bw_dist:gpu}). We observe that when the two GPUs are connected to the same switch, all systems exhibit a similar latency, between 3.7us and 5.7us. However, the location of the GPUs impacts both latency and goodput on all three systems. On Alps and LUMI, when the two GPUs are on different Dragonfly groups, the average latency increases by 28\% compared to the case where the two GPUs are under the same switch (from 4.33us to 5.56us on Alps), whereas on Leonardo it increases by 2x (from 2.03us to 4.23us). We observe a similar effect on the goodput. All three systems reach 95\% of theoretical peak bandwidth when the two GPUs are connected to the same switch. However, when the two GPUs are on different Dragonfly/Dragonfly+ groups, the average goodput decreases by 1\% on Alps and LUMI, and by 17\% on Leonardo (from 395 Gb/s to 328 Gb/s).

This is mostly caused by a large performance variability experienced on Leonardo when the two GPUs are not under the same switch. This is due \textit{network noise}~\cite{sc2019,htornetnoise,beni}, i.e., interference by other jobs sharing the same inter-node interconnection network. We analyze this in detail in Sec.~\ref{sec:congestion}. \textit{We observe that the 95th latency percentile on Leonardo increases to more than 8us when the two nodes are in different groups, with a maximum latency of 132us}. Similarly, we observed a minimum 216 Gb/s goodput.

\subsubsection{Host memory buffers} To provide a complete picture and isolate network performance from overheads related to GPU management, we report in Fig.~\ref{fig:lat_bw_dist:cpu} the same experiment, but transferring buffers allocated on host memory. %
We observe that latency on Leonardo is more than 3x smaller than on Alps and LUMI (1.02us vs. 3.66us for nodes connected to the same network switch). We attribute part of this difference to Slingshot relying on an Ethernet-based protocol, thus characterized by a slightly higher overhead compared to InfiniBand~\cite{Desensi2020} (e.g., due to larger headers~\cite{Desensi2020,10154243}). Although using the same network technology, Alps latency is slightly higher than LUMI's. However, Alps is not yet in production and optimizations across the entire stack are still ongoing. %

\obs{On Alps and LUMI, GPU's network location has a marginal impact on average performance (below 30\% for latency and 1\% for goodput). On the other hand, on Leonardo, the average latency increases by up to 2x when the GPUs are in different groups rather than under the same switch. Similarly, the average goodput decreases by 17\%. This is mainly due to network performance variability caused by network noise.}

\subsection{Alltoall}\label{sec:inter-node:coll:a2a}
We first analyze in Fig.~\ref{fig:collectives:scalability:a2a_gpus} the goodput of a 2 MiB alltoall when increasing the number of allocated GPUs. We consider an \emph{asymptotically} expected goodput, i.e., the expected goodput for a sufficiently large number of GPUs. This can be computed as the inter-node bandwidth available to each GPU (100 Gb/s on LUMI and Leonardo, and 200 Gb/s on Alps). This underestimates the actual goodput for a small number of GPUs since a larger fraction of communications happens on the intra-node rather than the inter-node network. The actual expected goodput can be easily computed by dividing the asymptotically expected goodput by the ratio of communications occurring on the inter-node network. However, we only report the asymptotic one to improve the readability of the plot.

\begin{figure}[htpb]
    \centering
    \includegraphics[width=.9\linewidth]{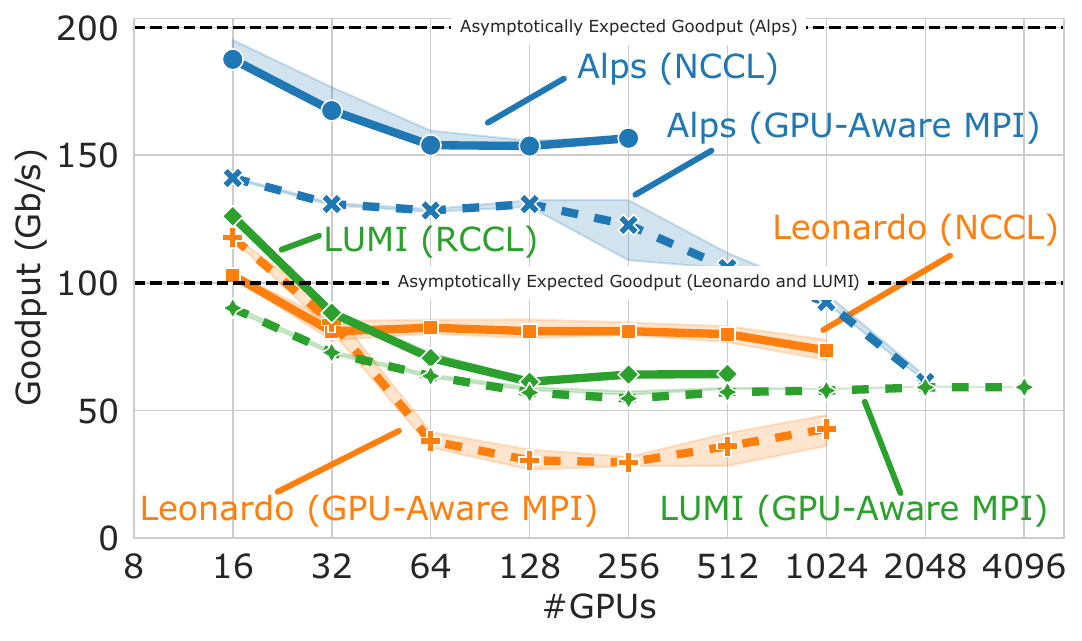}
    \caption{2 MiB alltoall scalability.}
    \label{fig:collectives:scalability:a2a_gpus}
\end{figure}

We did not run experiments up to \num{4096} GPUs on all the systems. Indeed, Leonardo measurements stop at \num{1024} GPUs since users can only run jobs using up to \num{256} nodes. On Alps, GPU-Aware measurements stop at \num{2048} GPUs since we have access to only \num{512} nodes at the time being. NCCL measurements stop at \num{256} GPUs since, for the alltoall, the benchmark gets stuck when running it on \num{512} GPUs and more (this occurs on both our benchmark and the official \texttt{nccl-tests}~\cite{nccl-test}). This is not the case for the allreduce collective (see Fig.~\ref{fig:collectives:scalability:ar_gpus}), and we can thus reasonably assume that the issue arises from the higher number of connections that must be kept active in the alltoall compared to the allreduce. It is worth remarking, however, that Alps is currently under deployment and still affected by some instability~\cite{santis}. Last, on LUMI, measurements for GPU-Aware MPI stop at \num{4096} GPUs, since users can allocate at most \num{512} nodes. RCCL measurements stop instead at \num{512} GPUs because, similarly to what we observed on Alps, the benchmark stalls when running it on \num{1024} GPUs and more (this happens on both our benchmark and on the official \texttt{rccl-tests}~\cite{rccl-test}). %

First, we observe that *CCL outperforms GPU-Aware MPI on all the systems. This happens because *CCL exploits the intra-node interconnect between GPUs more effectively, as discussed in Sec.~\ref{sec:intra-node-coll}. Indeed, the performance gap decreases when the number of GPUs increases, since the goodput becomes dominated by inter-node rather than intra-node performance. On Alps and Leonardo, *CCL achieves around 75\% efficiency up to \num{1024} GPUs, whereas on LUMI we observe a slightly lower efficiency.

\subsection{Allreduce}\label{sec:inter-node:coll:ar}
We perform a similar analysis in Fig.~\ref{fig:collectives:scalability:ar_gpus}, for a 1 GiB allreduce. For the allreduce, the memory occupancy is constant in the node count, rather than linear as for the alltoall, and we thus consider larger vectors. Consistent with what we observed for the alltoall, *CCL outperforms GPU-Aware MPI. As observed for the alltoall, this is due to *CCL being specifically optimized for the target architectures. On Leonardo, we observe an extremely low goodput for GPU-Aware MPI. As discussed in Sec.~\ref{sec:intra-node-coll:ar:measured}, this is due to Open MPI copying the buffer from the device to host memory and then running the allreduce on the host. We also observe a sharp drop in *CCL performance on Alps and LUMI from \num{256} to \num{512} GPUs. We exclude this is caused by a change in the allreduce algorithm, since the same drop happens when using the same algorithm on \num{256} and \num{512} GPUs, and the goodput steadily decreases between \num{256} and \num{512} GPUs, rather than dropping abruptly.

\begin{figure}[htpb]
    \centering
    \includegraphics[width=.9\linewidth]{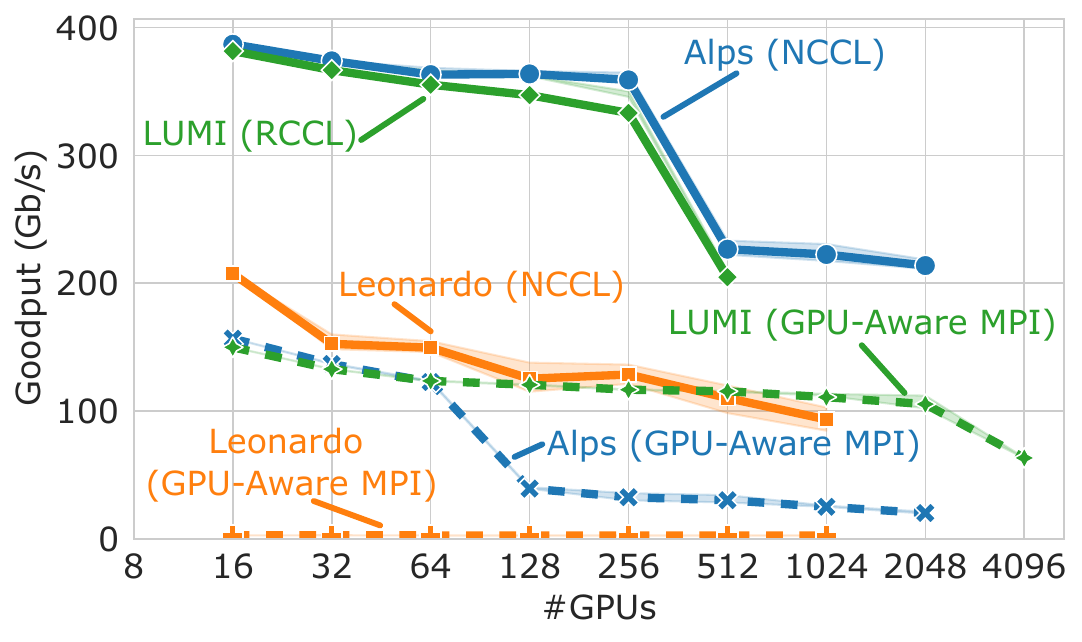}
    \caption{1 GiB allreduce scalability.}
    \label{fig:collectives:scalability:ar_gpus}
\end{figure}

\subsection{Comparison between MPI and *CCL}\label{sec:inter-node:coll:mpivccl}
Last, it is worth remarking that the performance gap between *CCL and MPI also depends on the vector size. Indeed, in Fig.~\ref{fig:collectives:scalability:heatmap}, we report the ratio between RCCL and GPU-Aware MPI for different node counts and vector sizes for alltoall and allreduce on LUMI. We can see that, whereas RCCL outperforms GPU-Aware MPI up to 4x on large vectors, for small collectives GPU-Aware MPI is characterized by up to 10x lower runtime. There is a sharp inversion of the trend around 32KiB, which we believe might be mitigated by tuning the allreduce algorithm selection. On Alps and Leonardo, instead, NCCL outperformed GPU-Aware MPI regardless of the message size and node count.

\begin{figure}[htpb]
\centering
\begin{subfigure}{0.49\linewidth}
    \includegraphics[width=1\linewidth]{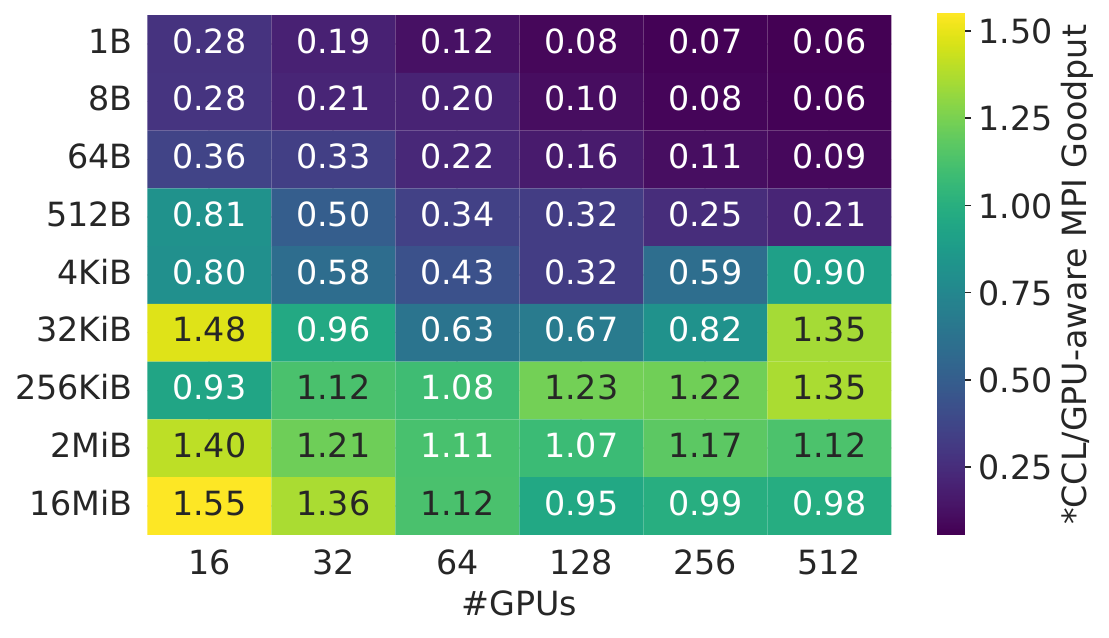}
    \caption{Alltoall}
    \label{fig:collectives:scalability:heatmap:a2a}
\end{subfigure}
\hfill
\begin{subfigure}{0.49\linewidth}
    \includegraphics[width=1\linewidth]{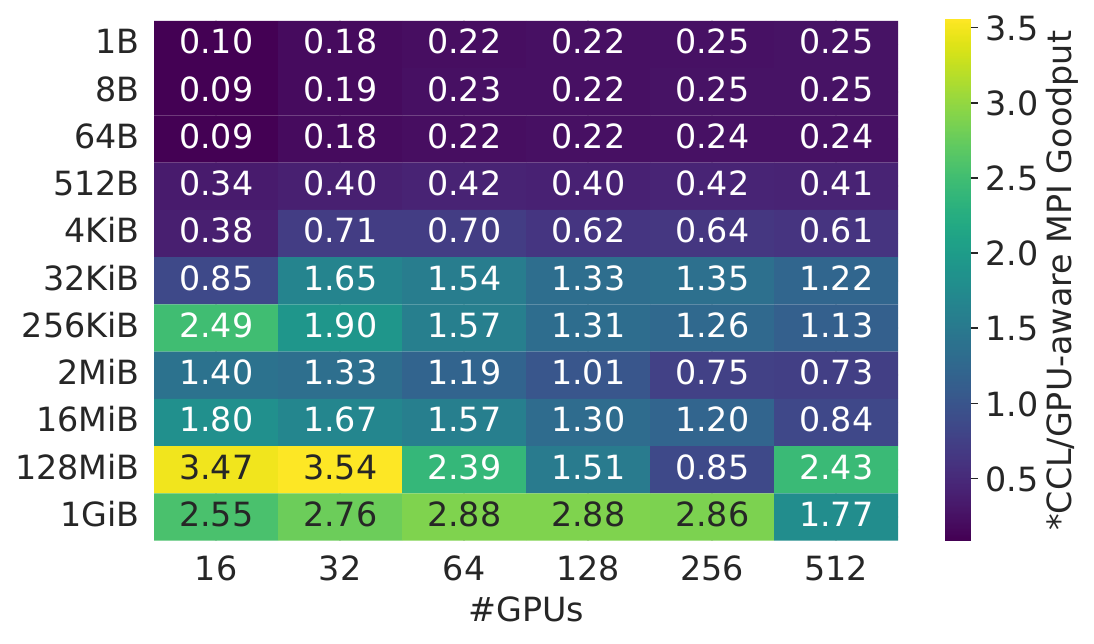}
    \caption{Allreduce}
    \label{fig:collectives:scalability:heatmap:ar}
\end{subfigure}
    \caption{Ratio between RCCL and GPU-Aware MPI goodput for different collective sizes and nodes count on LUMI.}
    \label{fig:collectives:scalability:heatmap}
\end{figure}

\obs{*CCL exploits the intra-node GPU-GPU interconnect more effectively than MPI, being specifically optimized for the target devices. Those advantages are more evident at smaller node counts and for larger transfers, for which the performance of intra-node communications has a higher weight on the overall performance. However, we experienced instability at large node counts for the alltoall on both NCCL and RCCL.}

\section{Network Congestion and Noise}\label{sec:congestion}
Sec.~\ref{sec:inter-node:location} shows that Leonardo is affected by network noise, which severely impacts performance when GPUs are not under the same network switch. In this section, we detail such impact and how it affects the scalability of collective operations. We have not performed a similar analysis on Alps and LUMI since, as shown in Sec.~\ref{sec:inter-node:location}, and also in previous works~\cite{Desensi2020,10.1145/3581784.3607089}, Slingshot is largely unaffected by network noise.

\subsection{Performance Isolation through Service Level Selection}\label{sec:leonardocongestion}
The variability observed in Sec.~\ref{sec:inter-node:location} largely comes from variable queueing delays experienced by packets when crossing the network (i.e., network noise). Indeed, we observed that variability only when the GPUs are not under the same switch. We thus try to reduce network performance variability by exploiting \emph{service levels}.

In InfiniBand, service levels can be used to mark the class of service of an application and are mapped to switch \emph{virtual lanes}. Each virtual lane is characterized by (logical) separate buffering and flow control. This means that applications share queues on network switches with other applications mapped to the same service level. Assuming a round-robin arbitration between the different virtual lanes, traffic forwarded on low-utilized service levels experiences lower queueing delays. 

Selecting a low-utilized service level can reduce the impact of network noise. On Leonardo, all the traffic is mapped by default on service level 0. To validate our hypothesis, we repeat the same experiment we ran in Fig.~\ref{fig:lat_bw_dist}, by selecting a \textit{service level} different from the default one\footnote{This can be done by setting the \texttt{NCCL\_IB\_SL} and \texttt{UCX\_IB\_SL} environment variables for NCCL and MPI, respectively.}. When switching to a service level different from the default one, we observed a significant reduction in performance variability, with a measured difference lower than $1\%$ between the minimum and maximum goodput for nodes in different groups (we do not explicitly show the results due to space constraints). It is worth remarking that, on Leonardo, adaptive routing is enabled on all service levels, and thus, the network noise reduction cannot be attributed to using or not using adaptive routing.

 \begin{figure}[htpb]
    \centering
    \includegraphics[width=1\linewidth]{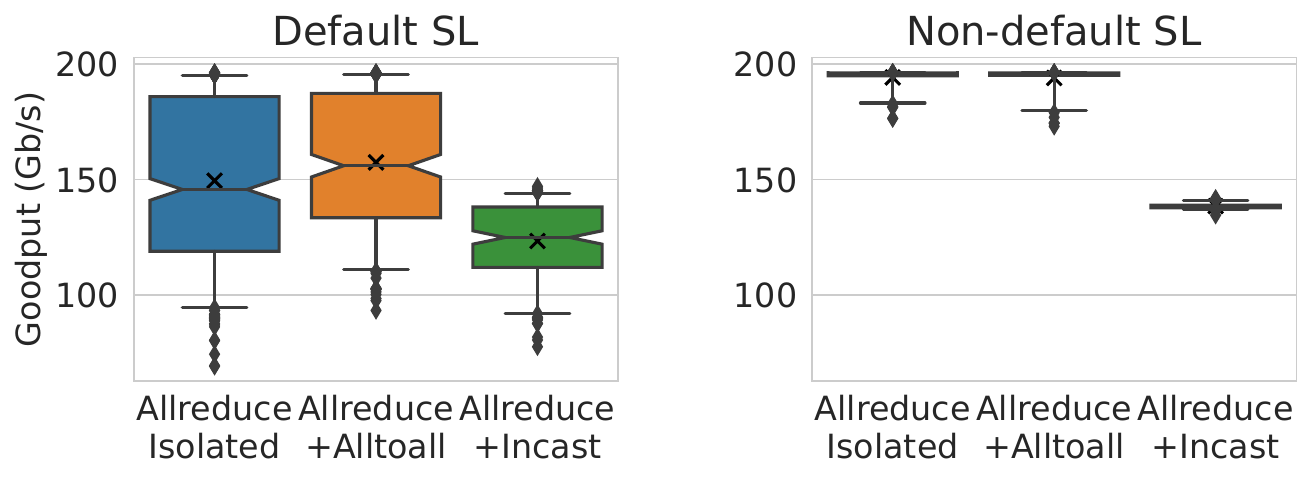}
    \caption{Impact of congestion on different service levels (SL) on Leonardo.}
    \label{fig:leonardo_sl_congestion}
\end{figure}

Although switching to a different service level mitigated the impact of network noise, it is important to note that this is only possible because, on Leonardo, all the traffic is mapped to the same service level by default. We would observe a similar performance variability if other applications run on the non-default service level. We demonstrate this by running an allreduce on 128 GPUs and, concurrently and on the same service level, another microbenchmark (running either an alltoall or an incast) on other 128 GPUs (benchmarks are allocated on nodes randomly). We repeat the same experiment both on the default service level and on a non-default service level. We show the result of our analysis in Fig.~\ref{fig:leonardo_sl_congestion}. We observe that, when the allreduce is run together with the incast, the goodput drops regardless of which service level we run the application on. We also ran the same test, but by allocating the nodes so to minimize the sharing of network switches between the two applications. In that case, we observed no performance impact of the incast on the allreduce (results are not shown in the figure due to space constraints). 

This demonstrates that network transfers on Leonardo are severely affected by network noise. Using a different service level only partially addresses the issue, and is strongly dependent on the number of jobs running on that service level. It is worth remarking that, to our knowledge, Leonardo is the only system deploying a Dragonfly+ topology at such a scale, and the routing algorithm might require further tuning to minimize the impact of network noise on performance.

\subsection{Noise Impact at Scale}
Studies on the impact of network noise on workload scalability have traditionally been carried out either through simulations~\cite{10.1145/3570609,htornetnoise}, or by injecting synthetic traffic~\cite{gpcnet,Desensi2020}. However, because on Leonardo all the traffic is mapped to the same service level, by analyzing the performance difference between the default and non-default service levels we have the unique opportunity to estimate the impact of real production network noise on multi-GPU workload scalability. Indeed, when the application runs on the default service level, it experiences the real production network noise, whereas when it runs on the non-default service level, it performs similarly to running on an empty system.

For this reason, we report in Fig.~\ref{fig:leonardo_cong_scalability} the goodput of a 2 MiB alltoall and a 1 GiB allreduce when running on the default and non-default service levels. We observe no differences when running on few GPUs, since only a few communications will occur between GPUs not connected to the same switch. When the number of GPUs increases, so does the number of inter-switch communications and the impact of congestion (i.e., the performance gap between the two service levels). Although, regardless of congestion, the performance decreases when increasing the number of GPUs (as described in Sec.~\ref{sec:inter-node}), we observe that, on \num{1024} GPUs, network noise causes an additional 20\% performance drop on alltoall, and a 50\% drop on the allreduce.
We want to remark that running on a non-default service level is only a temporary solution, possible because, at the time being, all the traffic runs by default on the same service level. Addressing this problem would thus require improvements in the adaptive routing algorithm. Moreover, on Leonardo Slurm is aware of which switch each node is connected to and to which Dragonfly+ group belongs, and can thus already optimize the job placement.

\obs{Network noise decreases the goodput of allreduce and alltoall up to 50\%.}

\begin{figure}[htpb]
\centering
\begin{subfigure}{0.49\linewidth}
    \includegraphics[width=1\linewidth]{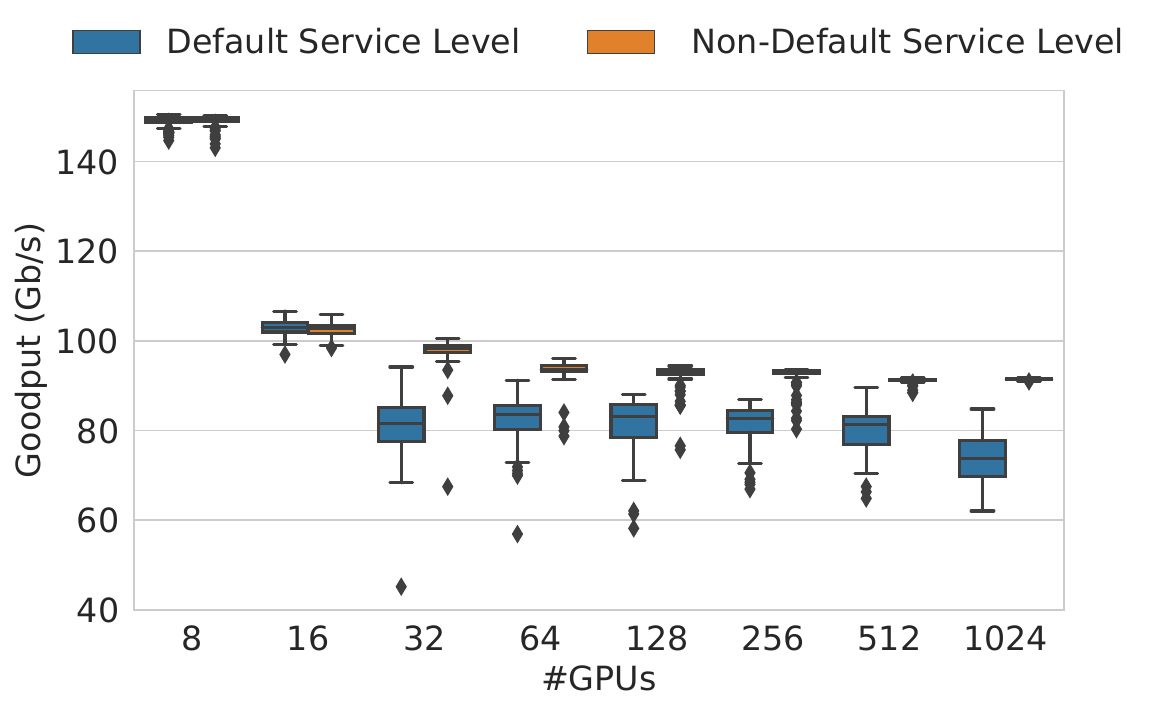}
    \caption{Alltoall}
\end{subfigure}
\hfill
\begin{subfigure}{0.49\linewidth}
    \includegraphics[width=1\linewidth]{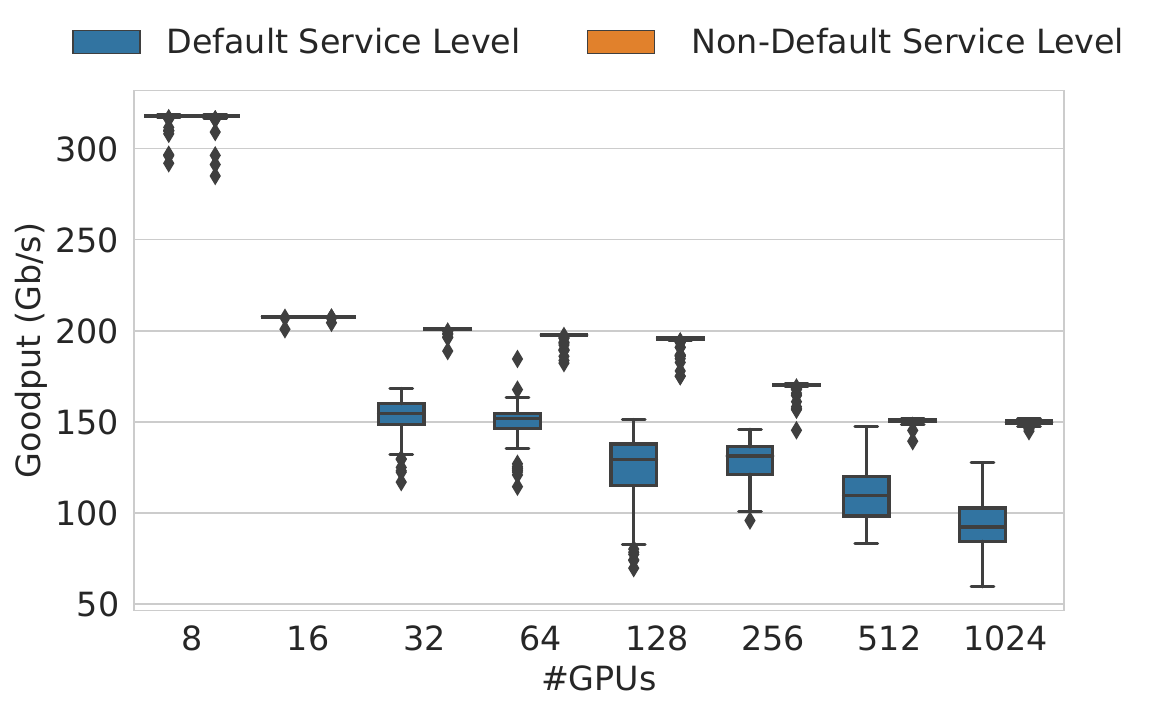}
    \caption{Allreduce}
\end{subfigure}
    \caption{Impact of congestion on scalability on Leonardo.}
    \label{fig:leonardo_cong_scalability}
\end{figure}

\section{State of the Art}

\subsection{Intra-node Interconnect} Different works analyzed the intra-node GPU-GPU interconnect. Pearson~\cite{pearson2023interconnect} characterizes the interconnect bandwidth heterogeneity within multi-GPU nodes using AMD MI250x GPUs, whereas Siefert et al.~\cite{10.1145/3624062.3624203} provide a detailed analysis of the intra-node GPU-GPU performance across various systems. However, neither paper analyzes inter-node performance. Moreover, while the former focuses on device-device copies without using any higher-level communication library, the latter only focuses on MPI. As we shown in Sec.~\ref{sec:intra-node-coll}, however, there are several tradeoffs to consider to determine the best communication library to use when moving data across GPUs on the same node, depending on the technology, transfer size, and communication pattern. 

Similarly, Atchley et al.~\cite{10.1145/3581784.3607089} characterize the \textit{Frontier} supercomputer, covering network, storage, and intra-node performance.
The analysis performed on the GPU interconnect, differently from this work, does not analyze the performance that can be obtained through different software solutions (ranging from device-device explicit copies to MPI). 
Moreover, although the paper shows full-scale results on an alltoall collective through MPI using GPCNet~\cite{gpcnet}, it does not specify whether the buffers are allocated on GPU memory (and thus GPU-Aware MPI is used) or on host memory. Because GPCNet allocates buffers on host memory~\cite{gpcnet-github}, we assume the tests did not use GPU-Aware MPI. Thus, on the GPU side, it only focuses on intra-node point-to-point transfers.

\subsection{Inter-node Interconnect} Other works extend the analysis to multiple nodes. Li et al.~\cite{Li2020} evaluate modern NVIDIA GPU interconnect technologies across different systems, including the Summit supercomputer~\cite{wells2016announcing}.
Khorassani et al.~\cite{khorassani2023high} compare different MPI implementations with RCCL for on multiple nodes of the Spock system, an early access cluster deployed with Slingshot and AMD MI100 GPUs. Both studies consider point-to-point and collective communications, but the analysis is limited to a few nodes (8 and 16, respectively). As we showed in Sec.~\ref{sec:inter-node} and Sec.~\ref{sec:congestion}, however, effects related to scalability and network noise are only visible at larger scales. 

Several works analyze network noise and interference between jobs at scale~\cite{10.1145/3581784.3607089,htornetnoise,Desensi2020,beni,10.1145/3570609,sc2019,staci2018,pollard2018}, often providing solutions to mitigate it. Most of these works, however, either simulate or synthetically generate noise. Differently from these works, in this paper we show the impact of noise induced by real production workloads.

\subsection{Other} 
Several benchmarks have been proposed to assess the performance of intra- and inter-node interconnection networks, including OSU~\cite{osu}, NCCL and RCCL test~\cite{nccl-test,rccl-test}, Tartan~\cite{Li2018}, and others~\cite{li2022}. The focus of this paper is, however, on the analysis of performance and scalability of multi-GPU supercomputers, rather than on the benchmarking itself. Last, some works analyze multi-GPU performance on several workloads including deep learning~\cite{tallent18,shi2018performance,7980078}, linear algebra~\cite{10155151}, computational physics~\cite{Gualtieri_Battista_Salvadore_Casciola_2023}, biology~\cite{OCETKIEWICZ2024109112}, data management~\cite{10.1145/3318464.3389705}, and others. Although this helps understand application scalability, it does not allow for evaluating the impact of the interconnect on the overall performance, nor to characterize and identify potential network bottlenecks.

\section{Discussion}

This study focuses on three specific supercomputers, but the benchmarks developed are general and applicable to other multi-GPU systems. Many conclusions, particularly those related to software, are broadly relevant. Tuning is expected to similarly impact other multi-GPU systems, and the performance differences between MPI and *CCL often stem from optimizations unique to one framework, suggesting similar behaviors elsewhere. We identify opportunities to improve GPU-aware MPI collective operations, especially allreduce, due to suboptimal host-GPU interactions during data aggregation.

Although none of the systems analyzed in this paper is based on fat tree networks, we expect most of our conclusions to hold regardless of the topology. The main exceptions are: i) very large fat tree systems may exhibit a slightly higher latency due to the greater diameter compared to Dragonfly/Dragonfly+; ii) the routing algorithm might differ on other topologies or technologies, and network noise levels may vary.

\section{Conclusions}
Understanding the performance of multi-GPU supercomputers is fundamental for identifying optimization opportunities. In this work, we focus on GPU-to-GPU communications, by thoroughly characterizing three supercomputers covering a significant fraction of the currently available HPC intra- and inter-node interconnect technology: Alps, Leonardo, and LUMI. 
Our analysis pinpoints several improvement opportunities, ranging from routing to communication libraries. 

First, the default software configuration of all three systems did not fully exploit their potential, requiring a non-negligible tuning effort to achieve good performance, both on a single node and at larger scales. Second, each communication library comes with its own set of optimizations and is thus more or less suitable according to the scenario. In general, we found *CCL to provide higher performance on collective operations, whereas GPU-Aware MPI performs better on point-to-point transfers. However, this depends on the system and the specific optimization implemented by MPI (e.g., on LUMI MPI outperforms RCCL on small collectives). Lastly, we showed that some existing HPC networks are still susceptible to network noise, which decreases performance at scale by up to 50\%.

We believe that our analysis can help users of large supercomputers to exploit those systems more efficiently. Furthermore, it provides valuable insights for systems and software designers as they address the challenges outlined herein.

\section*{Acknowledgment}
We thank Samuel Antao (AMD) for his feedback and support, and Kim McMahon (HPE Cray) for the enlightening discussion on MPICH performance on LUMI. We also thank CSCS for granting us early access to the Alps supercomputer. We acknowledge the CINECA award under the ISCRA initiative, for the availability of high performance computing resources and support. We acknowledge the EuroHPC Joint Undertaking for awarding this project access to the EuroHPC supercomputer LUMI, hosted by CSC (Finland) and the LUMI consortium through a EuroHPC Development Access call.
This work has been partially funded by Sapienza University under the SEED-2022 and
"Progetti Grandi 2023" funding schemes. Daniele De Sensi and Flavio Vella are members of \textit{Gruppo Nazionale Calcolo Scientifico - Istituto Nazionale di Alta Matematica} (GNCS-INdAM). Zebin Ren is funded by The Dutch Research Council (NWO) grant numbers OCENW.KLEIN.561.
The authors acknowledge financial support from \textit{ICSC – Centro Nazionale di Ricerca in High-Performance Computing, Big Data and Quantum Computing}, funded by European Union -- NextGenerationEU.
\bibliographystyle{IEEEtran}
\bibliography{main}

\twocolumn[%
{\begin{center}
\Huge
Appendix: Artifact Description/Artifact Evaluation        
\end{center}}
]

\appendixAD

\section{Overview of Contributions and Artifacts}

\subsection{Paper's Main Contributions}

\begin{description}
\item[$C_1$] Analysis of the intra-node interconnect in multi-GPU systems.
\item[$C_2$] Analysis of the inter-node interconnect in multi-GPU systems.
\item[$C_3$] Analysis of network noise impact on scalability in multi-GPU systems.
\end{description}

\subsection{Computational Artifacts}

\begin{description}
\item[$A_1$] https://zenodo.org/doi/10.5281/zenodo.13312325
\end{description}

\begin{center}
\begin{tabular}{rll}
\toprule
Artifact ID  &  Contributions &  Related \\
             &  Supported     &  Paper Elements \\
\midrule
$A_1$   &  $C_1$, $C_2$, $C_3$ & Fig. 3-13 \\
\bottomrule
\end{tabular}
\end{center}

\section{Artifact Identification}

\newartifact

\artrel

The artifact contains the benchmark code, the scripts used to run the benchmarks and collect the data, as well as the script to postprocess the data and generate the figures in the paper. The artifact also contains the data collected on the three analyzed systems.

\artexp

The results should be coherent with those presented in the paper when executed on systems with similar hardware and software environments. Some differences might be present due to differences in environments. For experiments involving multiple nodes, some small differences might be caused by job allocation and/or network configuration.

\arttime

On each system, we expect:
\begin{itemize}
\item Setup time: 5-10 minutes to download and compile the artifact.
\item Execution: 10 hours. This is the expected walltime. The actual compute hours depend on the number of GPUs allocated (e.g., more than 1000 compute hours might be needed for the tests involving 4096 GPUs). Most of the test can be scheduled with \texttt{sbatch}, but some of them require manual intervention to select a proper node allocation.
\item Analysis: 30 minutes. We provide scripts to produce the same plots shown in the paper.
\end{itemize}

\artin

\artinpart{Hardware}
Ideally, access to Alps, Leonardo, and LUMI is required. Alternatively, access to systems with a similar technology (e.g., Frontier).

\artinpart{Software}
Both the benchmarks code and the plotting scripts depend on the following Python packages: \texttt{seaborn}, \texttt{scipy}, \texttt{numpy}, \texttt{pandas}.
The benchmarks code also needs MPI and RCCL/NCCL, which should already be available on most systems. Different MPI and/or RCCL/NCCL versions might produce slightly different results, although we expect they will not alter the overall conclusions made in the paper.

\artinpart{Datasets / Inputs}
No datasets/inputs required.

\artinpart{Installation and Deployment}
The artifact does not rely on any specific compiler. However, the artifact heavily relies on Slurm as a workload manager.

\artcomp
The workflow consists of two main tasks: experiments execution, and plotting of the data. A \texttt{sbatch} script is provided to run the experiments needed for each of the figures in the paper. All the plots are produced by a single Python script.

\artout
The output of the benchmarks is automatically transformed from a human-readable format to a CSV file. All the CSV files are stored in the \texttt{data/} folder, together with a CSV file describing the setup of each experiment.

\newpage
\appendixAE

\arteval{1}
\artin
The framework has been configured so that each test will automatically load all the libraries and modules required. Additionally, the \texttt{./conf} folder contains one configuration file for each system, where the user can specify which additional modules to run and the paths of different libraries and executables required by the framework. Those paths have already been correctly set for the three systems analyzed in this work.

To plot the data, the following software is required: Python (at least 3.6.8), \texttt{numpy}, \texttt{pandas}, \texttt{seaborn}.

\artcomp
To compile the code, as a first step, the user should modify the \texttt{./conf.sh} file, and update the variable \texttt{BLINK\_SYSTEM} to the name of the target system (i.e., \texttt{alps}, \texttt{leonardo}, or \texttt{lumi}).
Then, by running \texttt{./compile.sh} all the needed code will be compiled. The script will compile the microbenchmarks targeting the transmission of GPU memory buffers (contained in the \texttt{./src/microbench-gpu} folder), as well as those targeting the transmission of host memory buffers (contained in the \texttt{./src/microbench} folder). The \texttt{./sbatch} folder contains all the Slurm scripts required to run the experiments we ran in the paper (divided into subfolders, one for each target system).

In the following, we describe the purpose of each script and the data they collect.
\begin{itemize}
    \item \texttt{./sbatch/SYSTEM/intra-node.sbatch} It collects intra-node performance data. This data is then used to plot Fig. 3, 5, and 6.
    \item \texttt{./sbatch/lumi/intra-node-gpu-pairs.sbatch} It collects the performance data for ping-pong between each pair of GPUs on a node. This data is then used to plot Fig. 4.
    \item \texttt{./sbatch/SYSTEM/two-nodes.sbatch} It collects the performance data for ping-pong between two GPUs on two different nodes. This data is then used to plot Fig. 7 and 8. For Fig. 8, the \texttt{ALLOCATION} variable must be set to either \texttt{same\_switch}, \texttt{diff\_switch} or \texttt{diff\_group} depending on whether the two nodes are under the same switch, under two different switches on the same group, or under two switches in two different groups. Unfortunately, checking the nodes location requires some manual effort from the user. 
    \begin{itemize}
    \item On Leonardo, this can be checked by analyzing the \texttt{topology.conf} file under \texttt{/var/spool/slurmd/conf-cache/}. The file contains, for each switch, the list of the nodes/NICs attached to that switch, as well as the list of switches within each dragonfly group (lines 12-33).
    \item On LUMI and Alps, on each node the \texttt{/etc/cray/xname} file contains the coordinates of that node. For example, on LUMI, it can be parsed as follows:
    \begin{itemize}
        \item x\#\#\#\# = network group / cabinet
        \item c\# = chassis (8 chassis / cabinet)
        \item s\# = blade (8 blades / chassis)
        \item b\# = board (2 boards / blade)
        \item n\# = node (1 node / board)
    \end{itemize}
    \end{itemize}
    \item \texttt{./sbatch/SYSTEM/many-nodes-coll.sbatch} It collects the performance for alltoall and allreduce on several nodes. This data is then used to plot Fig. 9, 10, and 13. This should be run several times, each with a different number of nodes/GPUs. 
    \item \texttt{./sbatch/leonardo/many-nodes-cong-sl.sbatch} It collects the data needed for Fig. 12.    
\end{itemize}

The framework saves the data under the \texttt{./data} folder. The \texttt{./data/description.csv} folder contains the metadata of each run. Please be aware that some of these experiments (e.g., those running on multiple nodes) might require thousands of compute hours.

\artout

All the paper plots can be generated by running the \texttt{./plots/plot.sh} script. We suggest doing that on a local machine. All the plots (with names matching their position in the paper) can be found under \texttt{./plots/paper}. The plots do not contain the annotations since those have manually added.

\end{document}